\documentclass[11pt,a4paper]{article}
\pdfoutput=1
\usepackage{bfss-jhep-compat}
\usepackage{jheppub}
\usepackage[T1]{fontenc}
\usepackage[utf8]{inputenc}
\usepackage{lmodern}
\usepackage{amsthm,mathtools}
\usepackage{microtype}
\usepackage{bookmark}
\hypersetup{pdftitle={D0-brane bound states in Matrix theory: nonperturbative bounds on matrix extent, noncommutativity and spatial shape},pdfauthor={Tetsuyuki Muramatsu}}
\numberwithin{equation}{section}
\newcommand{\Tr}{\operatorname{Tr}}
\newcommand{\tr}{\operatorname{tr}}
\newcommand{\Var}{\operatorname{Var}}
\newcommand{\Dom}{\operatorname{Dom}}
\newcommand{\R}{\mathbb R}
\newcommand{\HH}{\mathcal H}
\newcommand{\FF}{\mathcal F}
\newcommand{\dd}{\mathrm d}
\newcommand{\ii}{\mathrm i}
\newcommand{\CP}{C_{\mathrm P}}
\newcommand{\norm}[1]{\left\|#1\right\|}
\newcommand{\ev}[1]{\left\langle #1\right\rangle}
\allowdisplaybreaks[1]
\title{D0-brane bound states in Matrix theory:\\
nonperturbative bounds on matrix extent, noncommutativity and spatial shape}
\author[*]{\href{https://orcid.org/0009-0003-5107-8334}{Tetsuyuki Muramatsu}\note[*]{This work was carried out in a personal capacity. The views expressed are those of the author and do not necessarily reflect those of the affiliated organization.}}
\affiliation{Tokai National Higher Education and Research System,\\ Furo-cho, Chikusa-ku, Nagoya, 464-8601, Japan}
\emailAdd{muramatsu.tetsuyuki.m4@mail.f.thers.ac.jp}
\abstract{In Matrix theory, a single eleven-dimensional supergraviton is associated with a BPS bound state of D0-branes. Assuming the existence of a normalized physical BPS state, we derive nonperturbative bounds on its matrix extent, noncommutativity and spatial shape from supersymmetry in the BFSS model.

The probability of matrix configurations smaller than a fixed fraction of an explicit extent scale is exponentially suppressed in the square of the number \(N\) of D0-branes. In Yang--Mills variables, this scale for the trace-normalized extent is of order the cube root of the 't Hooft coupling. Relative noncommutativity, measured by a dimensionless commutator ratio, has an expectation bounded by a quantity of order the inverse of \(N\), and its probability of exceeding a sufficiently large fixed multiple of the inverse of \(N\) vanishes at large \(N\).

Concentration of the matrix extent along a fixed number of principal directions fewer than nine at large \(N\) requires the total extent relative to a corresponding reference scale to diverge in probability. It also requires relative noncommutativity to be suppressed more strongly than the inverse of \(N\), in every fixed finite moment of order at least one.

These bounds apply to the full quantum wave function, including off-diagonal matrix degrees of freedom, and constrain the quantum structure of D0-brane bound states and the possible large-\(N\) matrix geometries in Matrix theory.}
\keywords{M(atrix) Theories, M-Theory, Matrix Models, Nonperturbative Effects}
\begin{document}
\maketitle

\section{Introduction}
\label{sec:introduction}

Matrix theory was proposed as a description of uncompactified eleven-dimensional M-theory in the infinite momentum frame (IMF) through the large-\(N\) limit of the maximally supersymmetric \(U(N)\) matrix quantum mechanics of \(N\) D0-branes (the BFSS model) \cite{BFSS}. At finite \(N\), the BFSS model is conjectured to describe the sector of discrete light-cone quantization (DLCQ) of M-theory with \(N\) units of longitudinal momentum along a compact lightlike direction \cite{SusskindDLCQ}. This correspondence is supported by an appropriate weak-coupling limit of the D0-brane system, based on the relation between M-theory and type IIA string theory \cite{SenDLCQ,SeibergDLCQ}.

In Matrix theory, a normalizable BPS state of the interacting sector of the BFSS model is associated with a single eleven-dimensional supergraviton. After separating the \(U(1)\) center-of-mass sector, the relative degrees of freedom are described by nine traceless Hermitian \(N\times N\) matrices and their fermionic partners, with \(SU(N)\) gauge symmetry. When the matrices commute, their simultaneous eigenvalues can be interpreted as D0-brane positions, but general matrix configurations cannot be described by positions alone. To understand the quantum spatial structure of the bound state, we need to know how far the matrices extend, how strongly they fail to commute and how their extent is distributed among the nine transverse directions.

Previous studies have examined the symmetries of zero-energy states \cite{SS,HH}, the asymptotic behavior of the wave function for separated clusters \cite{FGHHY,HHasy,LY}, and matrix widths and correlation functions \cite{P,L,LZ}. More detailed information is available for the \(SU(2)\) bound state \cite{MSY,BLin}. Here we examine the probability distribution of full matrix configurations, including off-diagonal entries, to determine how supersymmetry constrains matrix extent, noncommutativity and spatial shape, and how these three properties are related.

We denote the total matrix extent by \(R\) and measure relative noncommutativity by a dimensionless ratio \(U\), the sum of squared commutator magnitudes divided by \(R^4\). Spatial shape is characterized by the distribution of squared extent along the principal directions. Even in a rotation-invariant quantum state, individual configurations can be anisotropic, so their shape can be examined without choosing a preferred orientation.

We assume the existence of a normalized physical BPS state satisfying the Gauss constraint and annihilated by all sixteen supercharges. Using the supersymmetric relation between kinetic energy, the bosonic commutator potential and the fermionic interaction, we obtain the following results for any such state, without assuming uniqueness or an explicit wave function.

The probability that the total extent \(R\) falls below a fixed fraction of an explicit scale is exponentially suppressed, with an exponent proportional to \(N^2-1\). In Yang--Mills variables, the corresponding scale for the trace-normalized extent is of order the cube root of the 't Hooft coupling. The expectation value of \(U\) is bounded by a quantity of order \(1/N\), and the probability that the rescaled ratio \(NU\) exceeds any fixed threshold greater than \(256/81\) vanishes at large \(N\).

For \(N\geq4\), we also derive an inequality relating matrix extent, noncommutativity and spatial shape. If the probability of configurations with appreciable extent outside a fixed number of principal directions fewer than nine vanishes at large \(N\), then \(R\) divided by a reference scale of the same large-\(N\) order as the small-extent scale must diverge in probability, and
\[
 \ev{(NU)^p}\longrightarrow0
 \qquad\text{for every fixed finite }p\geq1.
\]
Thus concentration along fewer spatial directions requires both an increase in relative extent and further suppression of relative noncommutativity.

These nonperturbative bounds follow from the BFSS model itself. Their interpretation in terms of the eleven-dimensional supergraviton uses the Matrix theory proposal, and they constrain the quantum matrix structure of the D0-brane bound state and the possible matrix geometries at large \(N\).

Section~\ref{sec:framework} introduces the BFSS model and the supersymmetric energy estimates, Section~\ref{sec:bounds} establishes the main results, and Section~\ref{sec:comparison} discusses their physical implications. Technical details and conditional extensions are given in the appendices.

\section{The BFSS model and physical zero-energy states}
\label{sec:framework}

We formulate the Hamiltonian and supersymmetry of the BFSS model,
then use them to derive an energy estimate for the probability
distribution of matrix configurations in a physical zero-energy state.

\subsection{Hamiltonian and physical zero-energy states}
\label{sec:hamiltonian}

For \(N\ge2\), we work in the interacting \(SU(N)\) sector of the
BFSS model after separating the center-of-mass motion. The bosonic
variables are nine traceless Hermitian matrices
\(X_i\coloneqq\sum_{a=1}^d X_{ai}T_a\), with real coefficients
\(X_{ai}\), \(i=1,\ldots,9\), and \(d\coloneqq N^2-1\).
The traceless Hermitian generators are normalized by
\[
 \Tr(T_aT_b)=\delta_{ab},\qquad
 [T_a,T_b]=\ii\sum_c f^{abc}T_c.
\]
We quantize the matrix components in temporal gauge \(A_0=0\)
with \(\hbar=1\), before imposing the residual Gauss constraint.
In the coordinate representation, the conjugate momenta are
\(P_{ai}=-\ii\partial_{ai}\), where
\(\partial_{ai}\coloneqq\partial/\partial X_{ai}\).
The fermions transform as real sixteen-component spinors of Spin(9)
and obey the canonical anticommutation relations
\[
 \{\theta_\alpha^a,\theta_\beta^b\}
   =\delta^{ab}\delta_{\alpha\beta},\qquad
 \alpha,\beta=1,\ldots,16.
\]
Let \(\Gamma_i\) be real symmetric \(16\times16\) matrices with
\(\{\Gamma_i,\Gamma_j\}=2\delta_{ij}\).

The Hamiltonian is
\[
 H\coloneqq K+V_B+V_F,\qquad
 K\coloneqq\frac12\sum_{a,i}P_{ai}^2=-\frac12\Delta .
\]
The bosonic potential and the fermionic interaction are
\begin{align}
 V_B&\coloneqq-\frac{g^2}{4}\sum_{i,j}\Tr[X_i,X_j]^2,
 &V_F&\coloneqq\frac{\ii g}{2}\sum_{a,b,c,i,\alpha,\beta}f^{abc}X_{ai}
             \theta_\alpha^b(\Gamma_i)_{\alpha\beta}\theta_\beta^c .
 \label{eq:potentials}
\end{align}
Here \(g>0\) is the coupling, and sums over transverse indices
\(i,j\) are ordered double sums.

We impose the Gauss constraint \(G^a\Psi=0\) on the states,
where \(G^a\) generates the residual \(SU(N)\) transformations
of both bosons and fermions. We impose this constraint without further
gauge fixing of the bosonic coordinates, so that all \(D\coloneqq9d\)
real coordinates are retained. The physical Hilbert space is
\begin{equation}
 \HH_N\coloneqq\bigl[L^2(\R^D)\otimes\FF_N\bigr]^{SU(N)},
 \label{eq:hilbert}
\end{equation}
where \(\FF_N\) carries the irreducible representation of the \(16d\)
real fermions. The sixteen supercharges are, with
\(\Gamma_{ij}\coloneqq[\Gamma_i,\Gamma_j]/2\),
\begin{align}
 q_\alpha&\coloneqq-\ii\sum_{a,i}C^\alpha_{ai}\partial_{ai}+V_\alpha(X),
 & C^\alpha_{ai}&\coloneqq\sum_\beta(\Gamma_i)_{\alpha\beta}\theta_\beta^a,
 \label{eq:supercharge}\\
 V_\alpha(X)&\coloneqq\frac g2\sum_{a,b,c,i,j,\beta}f^{abc}X_{bi}X_{cj}
                 (\Gamma_{ij})_{\alpha\beta}\theta_\beta^a .
 \label{eq:supercharge-potential}
\end{align}
On the physical Hilbert space, the supercharges satisfy the
supersymmetry algebra
\begin{equation}
 \{q_\alpha,q_\beta\}=2\delta_{\alpha\beta}H.
 \label{eq:model}
\end{equation}
In particular, \(q_\alpha^2=H\) for each \(\alpha\), so the
Hamiltonian is nonnegative on physical states.
Throughout, \(\Omega=\Omega_N\) denotes any normalized physical
BPS zero-energy state. For a wave function \(\psi\), we write
\(|\psi(X)|^2=\psi(X)^\dagger\psi(X)\), summing over all fermionic
components. For a vector of wave functions, the squared magnitude
also includes the sum over its vector components. The BPS condition, normalization and expectation value
of a function \(F\) are
\begin{equation}
 \begin{gathered}
 q_\alpha\Omega=0\quad(\alpha=1,\ldots,16),\qquad
 \int_{\R^D}\dd X\,|\Omega(X)|^2=1,\\
 \ev F\coloneqq\int_{\R^D}\dd X\,F(X)|\Omega(X)|^2.
 \end{gathered}
 \label{eq:bps-measure}
\end{equation}
Here \(|\Omega(X)|^2\) is the probability density of matrix configurations.
Our main bounds assume that such a state exists but require neither
its uniqueness nor its explicit wave function. Physical zero states
are known to be Spin(9) invariant \cite{SS,HH}; this result is needed
only for the supplementary fluctuation estimates.

To describe the distribution of matrix configurations, we use the
following measures of extent and noncommutativity:
\begin{equation}
 R^2\coloneqq\sum_i\Tr X_i^2=\sum_{a,i}X_{ai}^2,\qquad
 T\coloneqq-\sum_{i,j}\Tr[X_i,X_j]^2,\qquad
 U\coloneqq\frac{T}{R^4}.
 \label{eq:observables}
\end{equation}
Here \(R/\sqrt N\) measures the root-mean-square extent in the
normalized-trace convention, and the nonnegative quantity \(T\)
is related to the bosonic potential by \(V_B=g^2T/4\). The
dimensionless ratio \(U\) measures noncommutativity relative to the
total extent, independently of an overall rescaling of the matrices.
Its value at \(R=0\) is immaterial for all integrals. Both \(R\) and
\(U\) are invariant under transverse rotations and gauge transformations.

Appendix~\ref{app:operator} establishes self-adjoint realizations of
the supercharges and Hamiltonian on the physical Hilbert space, and
shows that the BPS wave function is smooth and locally bounded
throughout configuration space, including the origin.

\subsection{Energy estimates from supersymmetry}

The vanishing energy of a BPS state reflects a balance between the
positive kinetic and bosonic potential contributions and the
fermionic interaction, which can contribute negative energy.
We use this balance to constrain the probability distribution of
matrix configurations.

To probe particular regions of configuration space, multiply the BPS
wave function \(\Omega(X)\) by a real smooth gauge-invariant function
\(f(X)\) of compact support. This weights the probability density by
\(f^2\) without renormalizing the state. The resulting smooth physical
wave function \(f\Omega\) is generally not BPS, but its energy matrix
element can be calculated using supersymmetry. Since
\(q_\alpha^2=H\) and \(q_\alpha\Omega=0\),
\[
 (f\Omega,Hf\Omega)=\norm{q_\alpha(f\Omega)}^2
                  =\norm{[q_\alpha,f]\Omega}^2.
\]
The multiplication potential in the supercharge commutes with \(f\), so
\[
 [q_\alpha,f]\Omega=-\ii\sum_{a,i}C^\alpha_{ai}(\partial_{ai}f)\Omega .
\]
The canonical anticommutation relations give
\(\{C^\alpha_{ai},C^\alpha_{bj}\}=\delta_{ab}\delta_{ij}\)
for each \(\alpha\). Taking the squared norm yields
\begin{equation}
 h[f\Omega]
   =\norm{[q_\alpha,f]\Omega}^2
   =\frac12\ev{|\nabla f|^2}.
 \label{eq:multiplier}
\end{equation}
Here \(h[\psi]\coloneqq\norm{q_\alpha\psi}^2\) is the energy form,
defined on physical wave functions for which \(q_\alpha\psi\) is
square integrable, even if \(H\psi\) is not. Its value is independent
of \(\alpha\) (Appendix~\ref{app:operator}).
Equation~\eqref{eq:multiplier} shows that the energy cost of weighting
the BPS wave function is determined entirely by the gradient of the
weight and the original probability density.
On the other hand, the same energy can be computed from the three
terms of the Hamiltonian:
\[
 h[f\Omega]=K[f\Omega]+\ev{f^2V_B}
       +\int_{\R^D} f^2\Omega^\dagger V_F\Omega\,\dd X .
\]
Here \(K[\psi]=\tfrac12\int_{\R^D}\dd X\,|\nabla\psi|^2\)
is the kinetic contribution for the wave function \(\psi\), and
\(V_F\) acts on its fermionic components. Estimating these three
contributions and comparing them with \eqref{eq:multiplier} thus
constrains the BPS probability distribution without solving for
its wave function.

The uncertainty principle relates localization in the matrix
configuration space to kinetic energy. To study the probability of
configurations near the origin, we need a lower bound on kinetic
energy involving the inverse squared radius. For a smooth wave
function of compact support away from the origin, consider
\[
 0\le \int_{\R^D}
 \left|\nabla\psi+\frac{D-2}{2}\frac{X}{R^2}\psi\right|^2\dd X,
\]
where \(X\) denotes the vector of all \(D\) real matrix coefficients.
Expanding the square and integrating the cross term by parts, using
\(\nabla\cdot(X/R^2)=(D-2)/R^2\), gives
\[
 0\le 2K[\psi]-\frac{(D-2)^2}{4}
          \int_{\R^D}\dd X\,\frac{|\psi(X)|^2}{R^2}.
\]
The calculation applies to every fermionic component. We obtain
\begin{equation}
 K[\psi]\ge c_D\int_{\R^D}\dd X\,\frac{|\psi(X)|^2}{R^2},\qquad
 c_D\coloneqq\frac{(D-2)^2}{8}.
 \label{eq:hardy}
\end{equation}
This relation is known as the \(D\)-dimensional Hardy inequality
\cite{FS}. The extension to wave functions including the origin is justified in
Appendix~\ref{app:inverse-details}.
The nonnegative bosonic contribution is exactly
\(\ev{f^2V_B}=g^2\ev{f^2UR^4}/4\).
The fermionic interaction can contribute negative energy. To obtain
a lower bound on the total energy, we determine the lowest possible
fermionic energy at a given matrix configuration \(X\). Write
\[
 V_F(X)=\frac{\ii g}{2}\sum_{b,c,\alpha,\beta}
       \theta_\alpha^b A_{b\alpha,c\beta}(X)\theta_\beta^c,
 \qquad
 A_{b\alpha,c\beta}(X)=\sum_{a,i}f^{abc}X_{ai}(\Gamma_i)_{\alpha\beta}.
\]
The real antisymmetric \(16d\times16d\) matrix \(A\) can be brought
by an orthogonal transformation into two-dimensional blocks with
off-diagonal entries \(a_r\) and \(-a_r\), where \(a_r\ge0\).
Combining each pair of real fermions into a complex fermionic mode,
we obtain
\[
 V_F(X)=g\sum_{r=1}^{8d}a_r\left(n_r-\frac12\right),\qquad n_r=0,1.
\]
The lowest eigenvalue of \(V_F(X)\) on the full fermionic space is
\(-g\sum_r a_r/2\). To estimate this sum, we use the Cauchy--Schwarz
inequality and the trace of \(A^TA\):
\begin{equation*}
 \left(\sum_r a_r\right)^2\le8d\sum_r a_r^2,\qquad
 2\sum_r a_r^2=\Tr(A^TA)=32NR^2.
\end{equation*}
The trace is over the \(16d\) real fermion indices. The last equality
follows from \(\sum_{b,c}f^{abc}f^{ebc}=2N\delta_{ae}\) and
\(\Tr(\Gamma_i\Gamma_j)=16\delta_{ij}\). Thus the fermionic energy
is bounded from below by
\begin{equation}
 V_F(X)\ge -g C_NR\,I,\qquad
 C_N\coloneqq4\sqrt{2N(N^2-1)}.
 \label{eq:yukawa}
\end{equation}
Here \(I\) acts on the fermionic space. The spectrum at fixed \(X\)
is symmetric about zero, so every eigenvalue has absolute value at
most \(gC_NR\).
The lower bound holds in particular for physical wave functions.

Combining the three contributions
with the supersymmetric energy identity \eqref{eq:multiplier} gives
our basic inequality:
\begin{equation}
 c_D\ev{f^2R^{-2}}+\frac{g^2}{4}\ev{f^2UR^4}
 \le\frac12\ev{|\nabla f|^2}+gC_N\ev{f^2R}.
 \label{eq:master}
\end{equation}
This inequality constrains the extent to which the kinetic cost of
localization and the bosonic cost of noncommutativity can be offset
by the fermionic interaction, with the energy cost of the weight
accounted for by the gradient term. At small \(R\), the kinetic
lower bound grows as \(R^{-2}\), while the magnitude of the most
negative allowed fermionic energy is bounded by a quantity linear
in \(R\). At fixed extent, stronger noncommutativity increases the
bosonic potential. The inequality holds for real smooth gauge-invariant
functions \(f\) of compact support. Suitable choices of \(f\) turn
these competing energy contributions into quantitative probability
bounds on extent and noncommutativity.

\section{Probability bounds for D0-brane bound states}
\label{sec:bounds}

We first apply the supersymmetry inequality to small matrix
configurations and to their commutators relative to the total
magnitude. We then examine how these quantities constrain the
spatial shape of the configuration. All three sets of bounds
concern the same BPS wave function, with all matrix components retained.

\subsection{Probability of small matrix configurations}
\label{sec:inverse}

To estimate the probability of small matrix configurations, we bound
negative moments of \(R\). The inverse powers \(R^{-\nu}\) emphasize configurations with
small total extent. Choosing \(f=R^{-(\nu-2)/2}\) in
\eqref{eq:master}, so that \(f^2R^{-2}=R^{-\nu}\), gives a bound on
this moment together with a contribution from the bosonic potential.

For every real \(3\le\nu<D\), the choice of weight gives
\[
 |\nabla f|^2=\frac{(\nu-2)^2}{4}R^{-\nu}.
\]
Define the positive coefficient
\begin{equation}
 A_\nu\coloneqq c_D-\frac{(\nu-2)^2}{8}=\frac{(D-\nu)(D+\nu-4)}8>0.
 \label{eq:Anu}
\end{equation}
Substitution into \eqref{eq:master} gives
\begin{equation}
 A_\nu\ev{R^{-\nu}}+\frac{g^2}{4}\ev{UR^{6-\nu}}
 \le gC_N\ev{R^{3-\nu}}.
 \label{eq:mixed}
\end{equation}
The weight is singular at the origin and has no compact support.
Appendix~\ref{app:inverse-details} justifies the substitution with
radial cutoffs and proves that all terms are finite.
For \(\nu>3\), H\"older's inequality gives
\[
 \ev{R^{3-\nu}}\le\ev{R^{-\nu}}^{(\nu-3)/\nu}.
\]
Dropping the nonnegative commutator term and rearranging, we obtain
\begin{equation}
 \ev{R^{-\nu}}\le\left(\frac{gC_N}{A_\nu}\right)^{\nu/3}.
 \label{eq:inverse}
\end{equation}
At \(\nu=3\), the moment on the right of \eqref{eq:mixed} is one,
and the same conclusion follows directly.

At each fixed \(N\), smoothness of the wave function already
ensures that these inverse powers are integrable over the stated
range. The new information is the explicit bound, valid for every
normalized physical zero state, together with control of the
commutator contribution in the same inequality. To bound the probability of total extent at most \(R_0>0\), note that
\[
 \ev{R^{-\nu}}\ge\int_{R\le R_0}\dd X\,\frac{|\Omega(X)|^2}{R^\nu}
 \ge\frac{1}{R_0^\nu}\int_{R\le R_0}\dd X\,|\Omega(X)|^2.
\]
It follows from \eqref{eq:inverse} that
\[
 \int_{R\le R_0}\dd X\,|\Omega(X)|^2
 \le (R_0/r_{\nu,N})^\nu,
\]
where the reference scales are
\begin{equation}
 r_{\nu,N}\coloneqq\left(\frac{A_\nu}{gC_N}\right)^{1/3},\qquad
 \ell_N\coloneqq r_{D/2,N}=\left(\frac{D(3D-8)}{32gC_N}\right)^{1/3}.
 \label{eq:ell}
\end{equation}
Thus
\begin{equation}
 \int_{R\le\epsilon r_{\nu,N}}\dd X\,|\Omega(X)|^2
 \le\epsilon^\nu,\qquad 0<\epsilon<1.
 \label{eq:markov}
\end{equation}
Choosing \(\nu=D/2\), which is allowed for every \(N\ge2\), gives
\begin{equation}
 \int_{R\le\epsilon\ell_N}\dd X\,|\Omega(X)|^2
 \le\epsilon^{D/2}=\exp\!\left[-\frac92(N^2-1)|\log\epsilon|\right],
 \qquad 0<\epsilon<1.
 \label{eq:smallball}
\end{equation}

Equation~\eqref{eq:smallball} gives an upper bound, exponentially
small in \(N^2-1\), on the probability that the total matrix extent
of the D0-brane bound state is smaller than a fixed fraction of \(\ell_N\). The scale \(\ell_N\)
follows from the balance between the kinetic lower bound and the
fermionic interaction in the supersymmetric energy identity.
At large \(N\), the trace-normalized extent scale \(\ell_N/\sqrt N\) is of order
\(g^{-1/3}N^{1/3}\), in agreement with the scaling of previous
width estimates \cite{P,L}. The conversion to coordinates
with fixed Yang--Mills coupling is given in
Section~\ref{sec:comparison}.

The same finite-\(N\) inequality also bounds the bosonic commutator
term.

\subsection{Noncommutativity as the number of D0-branes increases}
\label{sec:commutator}

The inverse-radius estimate retains the bosonic interaction, which
allows us to bound noncommutativity in the BPS state. We first compare this quantum constraint with the range of \(U\)
for a single matrix configuration.

For nine Hermitian matrices of fixed nonzero total squared extent \(R^2\),
the squared commutator sum cannot exceed \(4R^4/3\). The sharp
configuration bound is
\begin{equation}
 0\le U\le\frac43
 \quad(N\ge2).
 \label{eq:ddvv}
\end{equation}
For example, \(X_i=t(\sigma_i\oplus0_{N-2})\), \(i=1,2,3\),
with all other \(X_i\) zero, has \(R^2=6t^2\) and \(T=48t^4\).
Thus even a two-dimensional color block can attain the maximum.
The general algebraic bound for Hermitian matrices is proved in
\cite{GXYYZ}.

\paragraph{Mean noncommutativity.}
Choosing \(\nu=6\) in \eqref{eq:mixed} cancels the overall radius
from the commutator term:
\begin{equation}
 A_6\ev{R^{-6}}+\frac{g^2}{4}\ev U\le gC_N\ev{R^{-3}},
 \qquad A_6=\frac{(D-6)(D+2)}8.
 \label{eq:coupled}
\end{equation}
Since \(\ev{R^{-6}}\ge\ev{R^{-3}}^2\), completing the square gives
\[
 \frac{g^2}{4}\ev U\le gC_N\ev{R^{-3}}-A_6\ev{R^{-3}}^2
 \le\frac{g^2C_N^2}{4A_6}.
\]
Substituting the coefficients, we obtain
\begin{equation}
 \ev U\le\frac{256N(N^2-1)}{(9N^2-15)(9N^2-7)}.
 \label{eq:U1}
\end{equation}
The algebraic bound \(\ev U\le4/3\) also holds.

The dynamical upper bound in \eqref{eq:U1} exceeds the algebraic
maximum at \(N=2\), but already gives \(512/407\approx1.26<4/3\approx1.33\) at \(N=3\); it decreases
monotonically for \(N\ge3\) and is asymptotic to \(256/(81N)\).
Thus the quantum bound is stronger than the maximum allowed by
matrix algebra alone, already at finite matrix size.

The dynamical coefficient in \eqref{eq:U1} is independent of \(g\),
as expected for a homogeneous observable of degree zero.
With normalized color trace \(\tr\coloneqq\Tr/N\), the corresponding quantity is
\begin{equation}
 NU=\frac{-\sum_{i,j}\tr[X_i,X_j]^2}
          {(\sum_i\tr X_i^2)^2}.
 \label{eq:normalized-U}
\end{equation}
The upper bound on the mean of \(U\) is of order \(1/N\). To control
the probability of large values of \(NU\), we need bounds on
higher moments.

\paragraph{Probability distribution of noncommutativity.}
To bound the second moment, we choose the weight
\(f=R^{-2}\sqrt U\), with a smooth regularization at \(U=0\).
The gradient term in \eqref{eq:master} now contains derivatives of
\(U\). Since \(U\) is invariant under a common rescaling of the
matrices, its gradient is orthogonal to the radial direction. With
\(X\cdot\nabla\coloneqq\sum_{a,i}X_{ai}\partial_{ai}\) denoting
the radial derivative multiplied by \(R\), we have the pointwise estimate
\begin{equation}
 X\cdot\nabla U=0,\qquad
 R^2|\nabla U|^2\le16U(2-U).
 \label{eq:shape-gradient}
\end{equation}
The first relation follows from scale invariance. The second is
obtained from the derivatives of the commutator sum and the bound
on the squared eigenvalue differences of a Hermitian matrix; the
calculation is given in Appendix~\ref{app:gradient-details}.

Using \eqref{eq:shape-gradient} in \eqref{eq:master}, we obtain,
for the same normalized physical BPS states,
\begin{equation}
 \ev{U^2}\le
 \frac{C_N^4}{c_D A_6}
 +\frac{16C_N^2}{A_6^2}
.
 \label{eq:U2}
\end{equation}
This upper bound is of order \(1/N^2\) at large \(N\).
The calculation uses the regularized weight
\(R^{-2}\sqrt{U+\eta}\), with \(\eta>0\). The derivative bound
controls its kinetic contribution, and the bosonic term in
\eqref{eq:master} gives the second moment of \(U\).
The explicit estimate and removal of the regularizations are given in
Appendix~\ref{app:gradient-details}.

The same choice of weight with higher powers of \(U\) gives bounds
on each fixed moment.
For every fixed real \(p\ge2\), Appendix~\ref{app:gradient-details}
gives an explicit finite-\(N\) bound on \(\ev{U^p}\), valid when
\(3p<D\). This condition is satisfied at sufficiently large \(N\)
for any fixed \(p\). The important result is that the same constant
controls all fixed moments at large \(N\): for every \(\delta>0\)
and every fixed \(p\ge2\), along any sequence of normalized physical
zero-energy states, at sufficiently large \(N\),
\begin{equation}
 \ev{(NU)^p}^{1/p}\le\frac{256}{81}+\delta.
 \label{eq:Lp}
\end{equation}
For any fixed threshold \(u>256/81\),
\[
 \int_{NU>u}\dd X\,|\Omega(X)|^2\le\frac{\ev{(NU)^p}}{u^p}.
\]
Choose \(\delta>0\) with \(256/81+\delta<u\). By \eqref{eq:Lp},
for sufficiently large \(N\) the probability is bounded by
\(((256/81+\delta)/u)^p\). This bound can be made arbitrarily small
by choosing a sufficiently large fixed \(p\) before taking \(N\) large.
Consequently, along the same sequence,
\begin{equation}
 \int_{NU>u}\dd X\,|\Omega(X)|^2\longrightarrow0,
 \qquad u>\frac{256}{81},\quad N\to\infty.
 \label{eq:threshold}
\end{equation}
Thus normalized noncommutativity exceeding any fixed threshold above
\(256/81\approx3.16\) has vanishing probability at large \(N\).

\paragraph{Finite and large matrix sizes.}
The high-moment bounds describe what is probable, rather than what
is algebraically possible. In fact, as shown in Appendix~\ref{app:limits}, at every fixed \(N\)
the wave function cannot vanish on an open region of configuration
space. Thus neighborhoods of configurations attaining the maximum
\(U=4/3\) still have positive probability. At fixed \(N\), increasing the moment order without bound therefore gives
\begin{equation}
 \lim_{p\to\infty}\ev{(NU)^p}^{1/p}=\frac{4N}{3}.
 \label{eq:order-limits}
\end{equation}
This diverges if \(N\) is subsequently sent to infinity. In contrast,
if the large-\(N\) limit is considered first at fixed \(p\), the upper
limit is bounded by \(256/81\) for every fixed \(p\), by \eqref{eq:Lp}.
The two orders of limits thus give different behavior. Our quantum
bound concerns the probability of large commutators, not their
algebraic possibility at a given matrix size.

\subsection{Spatial shape and the large-matrix-size limit}
\label{sec:spatial-shape}

We next ask how the total matrix extent is distributed among the
nine transverse directions. A Spin(9)-invariant quantum state can
contain configurations concentrated in subspaces of different
orientations. For a unit vector $n$ in the transverse space, the
squared extent along that direction is
$\Tr[(\sum_i n_iX_i)^2]=\sum_{i,j}n_i\Tr(X_iX_j)n_j$.
To measure the relative extent along each direction, introduce the
real symmetric matrix
\begin{equation}
 \mathsf S_{ij}\coloneqq\frac{\Tr(X_iX_j)}{\sum_l\Tr X_l^2}.
 \label{eq:spatial-gram}
\end{equation}
Its eigenvalues \(s_1\ge\cdots\ge s_9\) are the fractions of the
total squared extent along the principal directions and sum to one.
For \(N\ge4\), the nine vectors of color coefficients are linearly
independent except on a set of zero volume, so the eigenvalues
are positive with probability one. The fraction of extent squared
outside the \(k\) largest directions is
\begin{equation}
 \tau_k\coloneqq\sum_{j=k+1}^9s_j,\qquad 1\le k\le8.
 \label{eq:spatial-tail}
\end{equation}
Small \(\tau_k\) indicates concentration along \(k\) principal directions.
For example, a spherical membrane embedded in three transverse
directions has no extent in the other six directions, so $\tau_3=0$ for
its matrix embedding. The Gram eigenvalues measure extent in the
embedding space, not the intrinsic dimension of the membrane.

\paragraph{Spatial shape and kinetic energy.}
The kinetic bound \eqref{eq:hardy} grows when the total extent
becomes small. Even at fixed \(R\), however, localization near
configurations with small extent in individual directions has a
kinetic cost. The same square-completion method as in Section~\ref{sec:framework}
gives, for \(N\ge4\) and smooth test wave functions,
\[
 K[\psi]\ge\int\dd X\,|\psi(X)|^2
 \left(\frac{(N^2-11)^2}{8R^2}\sum_{j=1}^9\frac1{s_j}
 +\frac{968}{R^2}\right).
\]
The first term grows as any \(s_j\) approaches zero at fixed \(R\).
The derivation is given in Appendix~\ref{app:shape-proof}.
Using this kinetic bound in the supersymmetric energy estimate with
\(f=R^{-1/2}\) gives
\begin{equation}
 \begin{aligned}
 &\frac{(N^2-11)^2}{8}\ev{\frac1{R^3}\sum_{j=1}^9\frac1{s_j}}
 +\frac{g^2}{4}\ev{UR^3}
 +\frac{7743}{8}\ev{R^{-3}}\\
 &\qquad\le gC_N.
 \end{aligned}
 \label{eq:spatial-mixed}
\end{equation}
All three terms are finite; the cutoff argument is given in
Appendix~\ref{app:shape-proof}. This inequality relates spatial shape,
extent and noncommutativity. The kinetic cost of localization in
individual directions and the positive commutator contribution are
both constrained by the same fermionic energy bound.

\paragraph{Spatial shape and noncommutativity.}
Cauchy--Schwarz applied separately to the first \(k\) and last
\(9-k\) eigenvalues gives
\[
 \sum_{j=1}^9\frac1{s_j}\ge\frac{k^2}{1-\tau_k}
 +\frac{(9-k)^2}{\tau_k}.
\]
The first and second terms in \eqref{eq:spatial-mixed} contain
\(R^{-3}\) and \(R^3\), respectively. Applying the arithmetic--geometric
mean inequality pointwise to these contributions cancels the overall
extent and gives
\[
 \ev{\sqrt{\frac{NU}{\tau_k}}}\le
 \frac{16N\sqrt{N^2-1}}{(9-k)(N^2-11)}.
\]
Hence for \(\epsilon,u>0\),
\begin{equation}
 \int_{\substack{\tau_k\le\epsilon\\NU\ge u}}\dd X\,|\Omega(X)|^2
 \le\frac{16N\sqrt{N^2-1}}{(9-k)(N^2-11)}\sqrt{\frac\epsilon u}.
 \label{eq:spatial-joint-U}
\end{equation}
At large \(N\), the coefficient approaches \(16/(9-k)\).
The joint probability of strong concentration along a few directions
and appreciable normalized noncommutativity is thus constrained.

\paragraph{Spatial shape and total extent.}
The first term in \eqref{eq:spatial-mixed} also constrains the joint
probability of concentration and limited total extent. For
\(0<\epsilon<(9-k)/9\) and \(L>0\),
\begin{equation}
 \int_{\substack{\tau_k\le\epsilon\\R\le Lr_{3,N}}}\dd X\,|\Omega(X)|^2
 \le\frac{8A_3L^3}{(N^2-11)^2
 \left[\dfrac{k^2}{1-\epsilon}+\dfrac{(9-k)^2}{\epsilon}\right]}.
 \label{eq:spatial-joint-R}
\end{equation}
Here \(r_{3,N}\) is the reference scale in \eqref{eq:ell}.
The bracket in the denominator increases as \(\epsilon\) decreases
in the stated range. Thus concentration along a few directions
also constrains the probability of limited total extent.
Appendix~\ref{app:shape-proof} gives the derivation of both joint bounds.

\paragraph{Necessary conditions at large \(N\).}
For any sequence of normalized physical zero-energy
states and any fixed $k$, suppose that the matrix extent becomes
concentrated along the $k$ largest directions, in the sense that
\(\int_{\tau_k>\epsilon}\dd X\,|\Omega(X)|^2\to0\) for every
fixed $\epsilon>0$ as $N\to\infty$. Adding the probability of \(\tau_k>\epsilon\) to each joint bound,
then taking large \(N\) and finally small \(\epsilon\), shows that both
large normalized noncommutativity (above any fixed positive threshold)
and limited extent have vanishing probability. Combining this with
the fixed-moment bounds in Section~\ref{sec:commutator} gives
\begin{equation}
 \begin{gathered}
 \int_{R\le Lr_{3,N}}\dd X\,|\Omega(X)|^2\longrightarrow0
 \quad\text{for every fixed }L>0,\\
 \ev{(NU)^p}\longrightarrow0
 \quad\text{for every fixed }1\le p<\infty.
 \end{gathered}
 \label{eq:spatial-necessary}
\end{equation}

Here $r_{3,N}$ is the scale obtained from the bound on
$\ev{R^{-3}}$. The second condition also follows from
\eqref{eq:mixed} at $\nu=3$ once the growth of $R/r_{3,N}$ is known.
These are necessary conditions for spatial concentration; the bounds
do not determine whether it occurs or which $k$ is selected. The proof
is given in Appendix~\ref{app:shape-proof}.

\section{Physical interpretation and discussion}
\label{sec:comparison}

The bounds derived from supersymmetry relate matrix extent,
noncommutativity and spatial shape in D0-brane bound states. They
provide information about the quantum matrix structure underlying
the supergraviton description in Matrix theory. We discuss their
relation to previous results and their implications for understanding
this structure.

\subsection{Matrix extent and noncommutativity}

Previous studies have investigated matrix extent and noncommutativity
through expectation values of polynomial observables \cite{P,L,LZ}.
For a Spin(9)-invariant state with finite moments
$0<\ev{R^2}<\infty$ and $\ev T<\infty$, the ratio of expectation
values used in ref.~\cite{L} is related to our conventions by
\begin{equation}
 \frac{N\ev T}{\ev{R^2}^2}
 =-\frac89\frac{\ev{\tr[X_1,X_2]^2}}
 {\ev{\tr X_1^2}\ev{\tr X_2^2}} .
 \label{eq:gamma}
\end{equation}
Here the factor $8/9$ comes from the number of ordered pairs of
distinct transverse directions. In our approach the ratio is formed
for each configuration before averaging, giving
\(\ev{NU}=\ev{NT/R^4}\). This quantity and its higher moments are
well defined without assuming finite positive moments of $R$ or $T$,
and allow us to constrain the probability distribution of relative
noncommutativity directly.

To compare the extent scale with previous estimates, let $H_g(X)$
denote the Hamiltonian of Section~\ref{sec:hamiltonian} at coupling $g$, and let
$H_1(Y)$ denote the same Hamiltonian in matrix coordinates $Y_i$
with coupling set to one. The change of coordinates
\[
 X_i=g^{-1/3}Y_i,\qquad
 \frac{\partial}{\partial X_{ai}}=g^{1/3}
 \frac{\partial}{\partial Y_{ai}}
\]
extracts a common factor $g^{2/3}$ from the kinetic energy, the bosonic
potential and the fermionic interaction. With the derivatives
transformed as above, the Hamiltonian in the new coordinate
representation obeys the exact equality
\begin{equation}
 H_g(g^{-1/3}Y)=g^{2/3}H_1(Y).
 \label{eq:coupling-scaling}
\end{equation}
This change of coordinates can be implemented by a unitary dilation.
The factor $g^{-1/3}$ in $\ell_N$ is thus fixed by the scaling of
the Hamiltonian. In Yang--Mills variables $Z_i=gX_i$ with
't Hooft coupling $\lambda=g^2N$,
\begin{equation}
 \frac{\sqrt{\sum_i\tr Z_i^2}}{\lambda^{1/3}}
 =g^{1/3}N^{-5/6}R .
 \label{eq:units}
\end{equation}
At fixed $g$, the reference scale $\ell_N/\sqrt N$ is of order
$g^{-1/3}N^{1/3}$; at fixed $\lambda$, the corresponding scale for
the trace-normalized extent in $Z$ is of order $\lambda^{1/3}$.
This agrees with the scaling of previous width estimates \cite{P,L,LZ}.
Our result gives this scale a further quantum-mechanical meaning:
the probability of configurations below a fixed fraction of the
explicit reference scale is exponentially suppressed in $N^2$.
It extends the discussion of extent scaling to a nonperturbative
constraint on the full probability distribution.

\subsection{Relation to supermembrane matrix regularization}
\label{sec:classical-membrane}

Matrix regularization of the supermembrane \cite{dWHN} provides the
basis for the membrane description in Matrix theory \cite{BFSS}.
Let $x^i(\sigma)$ be the transverse coordinates of a fixed smooth
classical membrane with its center of mass removed. In matrix
regularization, a function $f$ on the membrane surface $\Sigma$ is
represented by an $N\times N$ matrix $Q_N(f)$. At large $N$,
\[
 \tr Q_N(f)\longrightarrow\int_\Sigma f\,\dd\mu_\Sigma,
\]
where $\dd\mu_\Sigma$ is the normalized area measure. The rescaled
commutator $-\ii[Q_N(f),Q_N(h)]/\kappa_N$ approaches the matrix
representation $Q_N(\{f,h\}_\Sigma)$ of the Poisson bracket, where
$\kappa_N$ is a regularization coefficient of order $1/N$.

Consider the matrices $Q_N(x^i)$, allowing an arbitrary common
nonzero rescaling, which cancels in $U$. For a membrane of nonzero
extent and a regularization with product and normalized-trace limits
sufficient to evaluate the squared commutators, eq.~\eqref{eq:normalized-U}
gives
\begin{equation}
 \lim_{N\to\infty}\frac{NU}{\kappa_N^2}
 =\frac{\displaystyle\int_\Sigma\sum_{i,j}
 \{x^i,x^j\}_\Sigma^2\,\dd\mu_\Sigma}
 {\displaystyle\left(\int_\Sigma\sum_i(x^i)^2\,\dd\mu_\Sigma\right)^2}.
 \label{eq:classical-membrane-U}
\end{equation}
Since $\kappa_N$ is of order $1/N$, eq.~\eqref{eq:classical-membrane-U}
implies $NU=O(N^{-2})$ for such a classical membrane sequence.

Our bounds instead apply to the full BPS probability distribution,
without specifying a classical membrane configuration. The algebraically
allowed two-dimensional Pauli color block has $U=4/3$ (so $NU=4N/3$),
whereas eq.~\eqref{eq:threshold} implies vanishing probability at large $N$
for configurations with $NU$ above any fixed threshold greater than $256/81$.
Our bounds control every fixed moment of $NU$. Under the concentration hypothesis
leading to eq.~\eqref{eq:spatial-necessary}, this boundedness is
strengthened to $\ev{(NU)^p}\to0$ for every fixed $1\le p<\infty$.
Determining the rate of this convergence requires further information
about the quantum state. This offers a way to investigate the relation between
the spatial structure of quantum matrix configurations and the
commutator scaling associated with membrane regularization.

In classical membrane regularization, increasing $N$ increases the
resolution of the membrane description. In the DLCQ interpretation
of the BFSS model, $N$ counts units of longitudinal momentum
\cite{SusskindDLCQ,SenDLCQ,SeibergDLCQ}. Understanding how these two
large-$N$ descriptions are related through the quantum dynamics is
an important question for the membrane interpretation of Matrix theory.

\subsection{Interplay between matrix extent, noncommutativity and spatial shape}
\label{sec:shape-interpretation}

A central result of this work is that supersymmetry imposes joint
constraints on matrix extent, noncommutativity and spatial shape.
The kinetic energy associated with quantum localization and the
bosonic potential associated with noncommutativity are both limited
by the fermionic energy estimate in eq.~\eqref{eq:spatial-mixed}.
Even at fixed total extent, concentrating the matrices along fewer
transverse directions increases the lower bound on the kinetic energy,
as the extent along the remaining directions decreases.

The joint probability bounds \eqref{eq:spatial-joint-U} and
\eqref{eq:spatial-joint-R} quantify this competition. For a fixed
number $k<9$ of principal directions, both bounds tend to zero
as the threshold $\epsilon$ tends to zero, uniformly at sufficiently large $N$ for
fixed thresholds on $NU$ and $R/r_{3,N}$. If such concentration
becomes dominant in a sequence of BPS states, eq.~\eqref{eq:spatial-necessary}
requires the total extent relative to $r_{3,N}$ to grow without bound
in probability and every fixed moment of $NU$ of order at least one
to vanish. The reference scales $r_{3,N}$ and $\ell_N$ have the
same large-$N$ scaling.

This has two implications for the spatial structure in Matrix theory.
First, the relative extent along different directions must be studied
alongside the large-$N$ scaling of the total extent. If a nonvanishing
fraction of the probability remains at a bounded multiple of the
reference scale, concentration along fewer than nine directions cannot
become dominant with probability tending to one. Second, such
concentration requires relative noncommutativity to be suppressed
more strongly than $1/N$. This provides a connection to the question
of how the commutator scaling of smooth membrane regularizations
might arise in quantum matrix dynamics.

The joint distribution of the three quantities is therefore a natural
object to study when investigating large-$N$ matrix geometry. Our
bounds provide nonperturbative necessary conditions for candidate
spatial structures of the full quantum state. Determining which
structure is realized, and whether it admits a continuum membrane
interpretation, is a question for the dynamics beyond these bounds.

\subsection{Outlook}

An important next step is to determine the scale of the typical
extent of a D0-brane bound state. Combining our probability bound
for small configurations with control of the probability at large
extent could make this possible. The relation between flat directions
of the potential and the continuous spectrum of supermembrane matrix
mechanics \cite{dWLN} is relevant to this problem. If the physical
BPS state is unique, supersymmetry and rotation symmetry also give
bounds on fluctuations across widely separated extent scales
(Appendices~\ref{app:projected-poincare}--\ref{app:log-radius}).
The ground-state problem on the valley regions \cite{BGR} and studies
of the asymptotic wave function \cite{FGHHY,HHasy,LY} are relevant to
understanding the existence, uniqueness and internal structure of a
normalized BPS state on the full configuration space.

A further question is the spatial structure of the matrices at large
$N$. The joint constraints derived here link extent, noncommutativity
and spatial shape through supersymmetry. Studying their joint
probability distribution can help clarify what spatial structures
arise in BFSS quantum mechanics and how they might be related to a
continuum quantum membrane description retaining the Poisson bracket.

Ultimately, Matrix theory aims to describe the physics of
eleven-dimensional M-theory through matrix quantum mechanics.
Connecting the internal matrix structure of D0-brane bound states to
supergraviton scattering would provide a way to investigate the relation
between matrix extent and a size inferred from scattering \cite{P}.
The longitudinal-momentum dependence of scattering amplitudes also
offers a connection to eleven-dimensional Lorentz symmetry, as in
soft-theorem arguments that require further assumptions on the amplitudes
and their large-$N$ limit \cite{HMSoft}. The nonperturbative information
obtained here from supersymmetry provides a starting point for these
questions.

\acknowledgments
OpenAI ChatGPT was used to assist with the exploration and development
of mathematical arguments, verification of derivations, editorial
organization, English drafting, reference and cross-reference checks,
and LaTeX preparation. The author takes full responsibility for the
scientific content of the manuscript.

\appendix
\section{Analytic details for the main bounds}
\label{app:moments}

The bounds in Section~\ref{sec:bounds} use weights that may be singular
at the origin or may extend to arbitrarily large matrix extent. We first
justify their use in the supersymmetric energy identity and then derive
the estimates on the commutator ratio. All expectations refer to the
normalized BPS wave function in Section~\ref{sec:framework}.
We use the two-sided fermion-energy bound stated after
\eqref{eq:yukawa}.

\subsection{Supercharges and regularity of the wave function}
\label{app:operator}

To apply the energy identity to weighted wave functions, we need a
self-adjoint realization of the supercharges and control of the wave
function near the origin. For fixed \(\alpha\), the canonical anticommutation relations imply
\[
 \{C^\alpha_{ai},C^\alpha_{bj}\}
   =\delta_{ab}\delta_{ij},\qquad
 \left(\sum_{a,i}C^\alpha_{ai}\xi_{ai}\right)^2=\frac{\sum_{a,i}\xi_{ai}^2}{2}I .
\]
Thus \(q_\alpha\) is a Dirac-type operator of constant propagation speed
\(1/\sqrt2\), with a smooth Hermitian matrix-valued potential. On every
ball this potential agrees with a bounded Hermitian truncation.
The truncated operator is essentially self-adjoint because it is a
bounded Hermitian perturbation of the free Dirac operator.
A general self-adjointness result for first-order operators with
bounded propagation speed \cite[Theorem 2.1]{Chernoff77} then gives
a self-adjoint closure of the original supercharge, initially defined
on smooth wave functions of compact support, including all fermion
components. We denote this initial space by
\(C_c^\infty(\R^D)\otimes\FF_N\).

Let \(P_G\) be compact-group averaging over \(SU(N)\). It commutes with
the supercharges and preserves compactness of support:
\begin{equation*}
 \operatorname{supp}(P_G\varphi)
 \subset SU(N)\cdot\operatorname{supp}\varphi .
\end{equation*}
The right-hand side is compact because the gauge group is compact.
Any physical wave function in the domain of a supercharge can
therefore be approximated by smooth gauge-invariant wave functions of
compact support, with convergence of both the wave function and its
supercharge image in norm. The space of these test wave functions,
\(\mathcal D_G\coloneqq P_G(C_c^\infty(\R^D)\otimes\FF_N)\),
is thus a common graph core for the supercharges. On this space
\(\norm{q_\alpha\psi}^2\) is independent of \(\alpha\).
Closing this nonnegative energy form gives $h$, with associated
Hamiltonian $H$ in \eqref{eq:model}. Since the self-adjoint supercharges
have the same square $H$ on the physical space, their common domain
consists of the wave functions of finite energy form, namely
\(\Dom(H^{1/2})\).

Denote the differential expression by
\(H_{\rm diff}\coloneqq-\Delta/2+V_B+V_F\).
For every test function in the full, unconstrained coordinate space,
gauge covariance gives
\begin{equation}
 (\Omega,H_{\rm diff}\varphi)
 =(\Omega,H_{\rm diff}P_G\varphi)
 =(q_\alpha\Omega,q_\alpha P_G\varphi)=0 .
 \label{eq:weak-full}
\end{equation}
The last identity uses the supersymmetry algebra on the physical space.
Since the first matrix element vanishes for every smooth test wave
function of compact support in the full coordinate space,
$H_{\rm diff}\Omega=0$ holds in the weak sense (with derivatives
understood through integration against test functions). The highest
derivatives in this equation are those of the Laplacian, and the
matrix potential is smooth. Local elliptic regularity therefore implies
that $\Omega$ is smooth and locally bounded, including at $X=0$.

\subsection{Inverse moments of the matrix extent}
\label{app:inverse-details}

We first extend the kinetic bound \eqref{eq:hardy} to wave functions
that need not vanish at the origin. The square-completion proof
applies directly to smooth functions supported away from the origin. For \(D>2\), it extends to smooth compactly
supported wave functions including the origin, by cutting out a
small ball and then taking its radius to zero. The gradient of the cutoff is bounded by a constant times
\(1/\epsilon\), and its support has volume of order \(\epsilon^D\).
For a smooth wave function the resulting squared-gradient error is
therefore bounded by a constant times \(\epsilon^{D-2}\) and tends to
zero. The inverse-square integral is locally finite for \(D>2\). Approximation in the
kinetic-energy domain and Fatou's lemma extend the bound to that domain.

Next we justify using the inverse powers of $R$ that lead to
\eqref{eq:mixed} and \eqref{eq:inverse}.
Set \(f(R)\coloneqq R^{-(\nu-2)/2}\). Since this function is singular
at the origin and is not compactly supported, we first multiply it
by smooth cutoffs that vanish near the origin and at sufficiently
large \(R\). Apply \eqref{eq:master} to
\(f_{\epsilon,L}\coloneqq\chi_\epsilon\chi_L f\).
For \(2\epsilon<L\), choose smooth radial cutoffs taking values between zero and one, with
\(\chi_\epsilon=0\) for \(R\le\epsilon\), \(\chi_\epsilon=1\) for
\(R\ge2\epsilon\), \(\chi_L=1\) for \(R\le L\), and \(\chi_L=0\)
for \(R\ge2L\). Their gradients are supported in
\((\epsilon,2\epsilon)\) and \((L,2L)\), respectively, and are bounded
by \(C/\epsilon\) and \(C/L\). Local smoothness and gauge invariance
of \(\Omega\) ensure that \(f_{\epsilon,L}\Omega\) belongs to the
physical test-function core \(\mathcal D_G\) defined in
Appendix~\ref{app:operator}.
At fixed \(N\), local boundedness of \(\Omega\) and \(\nu<D\)
give integrability of \(R^{-\nu}\) at the origin. At infinity
normalization suffices. The same reasoning controls \(R^{3-\nu}\)
because \(\nu\ge3\).

On the inner shell, $f^2|\nabla\chi_\epsilon|^2$ is bounded by a
constant times $\epsilon^{-\nu}$. The shell has volume of order
$\epsilon^D$, and $\Omega$ is locally bounded, so the squared-gradient
error is bounded by a constant times $\epsilon^{D-\nu}$ and vanishes.
On the outer shell, $f^2|\nabla\chi_L|^2$ is bounded by a constant
times $L^{-\nu}$; normalization of $\Omega$ then ensures that the
outer error also vanishes.
The cross terms vanish by Cauchy--Schwarz. For the nonnegative bosonic
potential term use Fatou's lemma; all other terms pass to the limit
by the preceding integrability bounds. Since
\[
 |\nabla f|^2=\frac{(\nu-2)^2}{4}R^{-\nu},
\]
the limiting inequality is \eqref{eq:mixed}.
The same argument establishes finiteness of the positive commutator
term in \eqref{eq:mixed}.

\subsection{Higher moments of noncommutativity}
\label{app:gradient-details}

The higher moments of $U$ involve derivatives of the commutator ratio.
To bound them, first differentiate the numerator $T$. Put
\(B_i\coloneqq\sum_j[X_j,[X_j,X_i]]\).
Then \(\nabla_iT=4B_i\) and \(\sum_i\Tr(X_iB_i)=T\).
For a Hermitian matrix \(Y\), diagonalizing \(X_j\) with eigenvalues
\(\lambda_a\) gives
\[
 -\Tr[X_j,Y]^2=\sum_{a,b}(\lambda_a-\lambda_b)^2|Y_{ab}|^2
 \le 2\Tr(X_j^2)\Tr(Y^2).
\]
Here the diagonal terms vanish, and for $a\ne b$,
$(\lambda_a-\lambda_b)^2\le2(\lambda_a^2+\lambda_b^2)
\le2\Tr(X_j^2)$.
The linear map \(\mathcal L_XY\coloneqq\sum_j[X_j,[X_j,Y]]\)
is symmetric with respect to \(\Tr(YZ)\). The preceding estimate
shows that its eigenvalues lie between zero and \(2R^2\).
Expanding \(X_i\) in its eigenmatrices therefore gives
\[
 \sum_i\Tr(B_i^2)\le2R^2\sum_i\Tr(X_iB_i)=2R^2T.
\]
The derivative of $U=T/R^4$ is orthogonal to the radial direction.
Using $X\cdot\nabla T=4T$ gives
\[
 R^2|\nabla U|^2
 =\frac{|\nabla T|^2}{R^6}-16U^2
 \le 32U-16U^2,
\]
which proves \eqref{eq:shape-gradient}.

Related derivative bounds for homogeneous polynomials, including
references to earlier work, can be found in \cite{vCS,vCSerr}.

For the second-moment bound \eqref{eq:U2}, use \(f_\eta\coloneqq R^{-2}\sqrt{U+\eta}\), \(\eta>0\), with inner and
outer radial cutoffs. Homogeneity makes the radial and angular
derivatives orthogonal, and \eqref{eq:shape-gradient} gives
\[
 |\nabla f_\eta|^2
 =R^{-6}\left(4(U+\eta)
       +\frac{R^2|\nabla U|^2}{4(U+\eta)}\right)
 \le(8+4\eta)R^{-6}.
\]
Remove the radial cutoffs as in Appendix~\ref{app:inverse-details} and
then let \(\eta\downarrow0\). The boundedness of \(U\) and
\(\ev{R^{-6}}<\infty\) justify these limits. Equation~\eqref{eq:master}
becomes
\[
 c_D\ev{UR^{-6}}+\frac{g^2}{4}\ev{U^2}
 \le gC_N\ev{UR^{-3}}+4\ev{R^{-6}}.
\]
Use
\(\ev{UR^{-3}}^2\le\ev{UR^{-6}}\ev U\),
complete the square, and insert
\(\ev U\le C_N^2/A_6\) and
\(\ev{R^{-6}}\le(gC_N/A_6)^2\).

To obtain the large-$N$ probability bound, it is enough to control
every fixed moment of $NU$ as $N$ increases. Here we give an explicit
finite-$N$ estimate for that purpose. Fix $p\ge2$ with $3p<D$, and
write $m=p-1$ to simplify the powers in the calculation. Define
\begin{equation}
 a_m\coloneqq c_D-2+2m^2,\qquad
 b_p\coloneqq\frac{C_N^2}{a_m},\qquad
 d_p\coloneqq\frac{16m^2C_N^2}{A_{3p}^2}.
 \label{eq:pconstants}
\end{equation}
The finite-\(N\) estimate is
\begin{equation}
 \ev{U^p}^{1/p}\le
 \frac{b_p+\sqrt{b_p^2+4d_p}}2 .
 \label{eq:Up}
\end{equation}

Use
\[
 f_\eta\coloneqq R^{-2}(U+\eta)^{m/2},\qquad 0<\eta\le1,
\]
with inner and outer radial cutoffs. First remove the radial
cutoffs at fixed \(\eta\), then let \(\eta\downarrow0\).
For \(m\ge1\), the squared gradient is dominated by a constant,
depending on the fixed \(m\), times \(R^{-6}\): when \(1\le m<2\),
the factor
\(U(U+\eta)^{m-2}\le(U+\eta)^{m-1}\)
controls the possible singularity at \(U=0\).
Boundedness of \(U\) and \eqref{eq:inverse} justify the
limits. At \(m=1\), \(U^{m-1}\) in the ensuing upper bound is
understood to be one, including at \(U=0\).

After the radial cutoffs are removed, radial-angular orthogonality
and \eqref{eq:shape-gradient} give
\begin{equation*}
 \frac12|\nabla f_\eta|^2
 \le R^{-6}\bigl[2(U+\eta)^m
       +2m^2(U+\eta)^{m-2}U(2-U)\bigr].
\end{equation*}
Apply \eqref{eq:master} to $f_\eta$ with this upper bound on its
kinetic contribution. As $\eta\downarrow0$, the expression in
brackets tends, for $U>0$, to $(2-2m^2)U^m+4m^2U^{m-1}$.
At $U=0$ its limit is zero, also bounded by this expression with
the convention for $m=1$ stated above. The domination by a constant
times $R^{-6}$ justifies taking the limit inside the expectation.
The resulting inequality is
\begin{equation}
 a_m\ev{U^mR^{-6}}+\frac{g^2}{4}\ev{U^{m+1}}
 \le gC_N\ev{U^mR^{-3}}
       +4m^2\ev{U^{m-1}R^{-6}}.
 \label{eq:power-master}
\end{equation}
Weighted Cauchy--Schwarz bounds the first term on the right by
\(gC_N\ev{U^mR^{-6}}^{1/2}\ev{U^m}^{1/2}\).
Completing the square therefore gives
\begin{equation}
 \ev{U^p}\le b_p\ev{U^{p-1}}
       +\frac{16m^2}{g^2}\ev{U^{p-2}R^{-6}} .
 \label{eq:moment-recursion}
\end{equation}
H\"older's inequality and the inverse-radius estimate at \(\nu=3p\)
give
\begin{align*}
 \ev{U^{p-1}}&\le \ev{U^p}^{(p-1)/p},\\
 \ev{U^{p-2}R^{-6}}
 &\le \ev{U^p}^{(p-2)/p}\ev{R^{-3p}}^{2/p}
 \le \ev{U^p}^{(p-2)/p}
             \left(\frac{gC_N}{A_{3p}}\right)^2 .
\end{align*}
The second line includes \(p=2\), with its zero power interpreted
as one. If \(\ev{U^p}=0\), the claim is immediate. Otherwise,
substituting these bounds into \eqref{eq:moment-recursion} and dividing
by \(\ev{U^p}^{(p-2)/p}\) gives
\[
 \ev{U^p}^{2/p}\le b_p\ev{U^p}^{1/p}+d_p.
\]
Solving this quadratic inequality for the nonnegative quantity
\(\ev{U^p}^{1/p}\) proves \eqref{eq:Up}. For every fixed \(p\),
\begin{equation*}
 Nb_p\longrightarrow\frac{256}{81},
 \qquad N^2d_p\longrightarrow0,
\end{equation*}
which proves \eqref{eq:Lp}. Here $p$ is held fixed throughout the large-$N$ limit.

\subsection{Order of the large-\texorpdfstring{$N$}{N} and large-moment limits}
\label{app:limits}

The upper-tail bound \eqref{eq:threshold} concerns increasing $N$.
At fixed $N$, we can instead ask whether configurations near the
algebraic upper bound $U=4/3$ carry nonzero probability.
The weak equation \eqref{eq:weak-full} has finitely many components,
a common Laplace principal part and a matrix potential bounded on
any ball $B$. On that ball it implies
\[
 |\Delta\Omega|\le C_{N,B}|\Omega|.
\]
Unique continuation for such finite systems
\cite[p.~248, Remark 3]{Aronszajn} implies that a nonzero normalized
solution cannot vanish throughout any open region. By continuity,
each nonempty open region has positive BPS probability.

The Pauli configuration attaining $U=4/3$ in \eqref{eq:ddvv} lies
away from $R=0$. For any $0<\delta<4/3$, continuity of $U$ gives an
open neighborhood of this configuration on which $U>4/3-\delta$.
Consequently, at fixed $N$,
\[
 N(4/3-\delta)
 \left(\int_{U>4/3-\delta}\dd X\,|\Omega(X)|^2\right)^{1/p}
 \le \ev{(NU)^p}^{1/p}\le 4N/3.
\]
The integral is strictly positive, so we can first take $p\to\infty$
and then $\delta\to0$. This yields
\begin{equation}
 \lim_{p\to\infty}\ev{(NU)^p}^{1/p}=\frac{4N}{3}.
 \label{eq:essential-sup}
\end{equation}
This is the fixed-$N$ limit in \eqref{eq:order-limits}. The probabilities
of these neighborhoods can decrease as $N$ increases, consistently
with \eqref{eq:threshold}. The argument holds for each normalized
BPS state separately.

\section{Derivation of the spatial-shape constraints}
\label{app:shape-proof}

In Section~\ref{sec:framework} we bounded the kinetic energy associated
with localization near small matrix configurations. Here we extend this
estimate to localization in individual spatial directions, derive the
mixed bound \eqref{eq:spatial-mixed}, and obtain the joint probability
bounds of Section~\ref{sec:spatial-shape}. We take $N\ge4$ throughout.

\subsection{Kinetic energy and spatial localization}

Represent the nine matrices by the real $d\times9$ coefficient
matrix $x=(X_{ai})$, with $d=N^2-1$, $\mathsf W=x^Tx$ and
$R^2=\Tr_9\mathsf W$. Here $\Tr_9$ denotes the ordinary trace over
the nine spatial directions. If $s_j$ are the eigenvalues of the matrix
$\mathsf S=\mathsf W/R^2$ in \eqref{eq:spatial-gram}, then
\[
 \Tr_9(\mathsf W^{-1})=\frac1{R^2}\sum_{j=1}^9\frac1{s_j}.
\]
This quantity increases when an individual principal extent becomes
small even at fixed $R$. It therefore provides a way to extend the
radial kinetic estimate to localization in individual directions.

As in the radial calculation, for a real vector field $V$ we expand
 the nonnegative integral and integrate the cross term by parts:
\begin{align*}
 0&\le\int\dd X\,|\nabla\eta+V\eta|^2\\
 &=2K[\eta]+\int\dd X\,(|V|^2-\nabla\cdot V)|\eta|^2.
\end{align*}
Thus $K[\eta]\ge\frac12\int\dd X\,(\nabla\cdot V-|V|^2)|\eta|^2$.
To include both the individual principal extents and the overall extent,
consider, where $\mathsf W$ is invertible,
\[
 V=\alpha x\mathsf W^{-1}+\beta\frac{x}{R^2}.
\]
Write $q=d-10=N^2-11>0$ for the calculation. Differentiation with
respect to the $D=9d$ real coefficients, and contraction over these
coefficients, give
\begin{align*}
 \nabla\cdot(x\mathsf W^{-1})&=q\Tr_9\mathsf W^{-1},&
 |x\mathsf W^{-1}|^2&=\Tr_9\mathsf W^{-1},\\
 \nabla\cdot\frac{x}{R^2}&=\frac{D-2}{R^2},&
 (x\mathsf W^{-1})\cdot\frac{x}{R^2}&=\frac9{R^2}.
\end{align*}
Together with $|x/R^2|^2=R^{-2}$, these relations give
\begin{align*}
 \frac{\nabla\cdot V-|V|^2}{2}
 ={}&\frac{\alpha(q-\alpha)}2\Tr_9\mathsf W^{-1}\\
 &+\frac{\beta(D-2-18\alpha-\beta)}{2R^2}.
\end{align*}
The coefficient of the inverse-matrix term is largest at $\alpha=q/2$.
With this choice, $D-2-18\alpha=88$, so the radial coefficient is
largest at $\beta=44$ and equals $968$. This gives the vector field
$qx\mathsf W^{-1}/2+44x/R^2$.

To apply the bound also to wave functions which do not vanish near
rank-deficient configurations or the origin, we regularize this vector
field. For $\varepsilon,\zeta>0$, set
\begin{equation}
 B_\varepsilon=(\mathsf W+\varepsilon I_9)^{-1},\quad
 Q=R^2+\zeta,\quad
 V_{\varepsilon,\zeta}=\frac q2xB_\varepsilon+\frac{44x}{Q}.
 \label{eq:appE-vector}
\end{equation}
Using $\dd B_\varepsilon=-B_\varepsilon(\dd\mathsf W)B_\varepsilon$
and $\mathsf WB_\varepsilon=I_9-\varepsilon B_\varepsilon$, we obtain
\begin{align*}
 \nabla\cdot(xB_\varepsilon)
 &=q\Tr_9B_\varepsilon+\varepsilon\bigl[\Tr_9(B_\varepsilon^2)
                         +(\Tr_9B_\varepsilon)^2\bigr],\\
 |xB_\varepsilon|^2&=\Tr_9B_\varepsilon-\varepsilon\Tr_9(B_\varepsilon^2).
\end{align*}
For the radial term and the cross term,
\begin{align*}
 \nabla\cdot\frac{x}{Q}&=\frac{D-2}{Q}+\frac{2\zeta}{Q^2},\\
 \left|\frac{x}{Q}\right|^2&=\frac1Q-\frac{\zeta}{Q^2},\\
 (xB_\varepsilon)\cdot\frac{x}{Q}
 &=\frac{9-\varepsilon\Tr_9B_\varepsilon}{Q}.
\end{align*}
Combining these expressions gives the exact identity
\begin{align}
 \frac{\nabla\cdot V_{\varepsilon,\zeta}
           -|V_{\varepsilon,\zeta}|^2}{2}
 ={}&\frac{q^2}{8}\Tr_9B_\varepsilon+\frac{968}{Q}
 \notag\\
 &+\varepsilon\left[\frac{q(q+2)}8\Tr_9 B_\varepsilon^2
  +\frac q4(\Tr_9B_\varepsilon)^2
  +\frac{22q}{Q}\Tr_9B_\varepsilon\right]
  +\frac{1012\zeta}{Q^2}.
 \label{eq:appE-gsr}
\end{align}
All terms proportional to $\varepsilon$ or $\zeta$ are nonnegative
since $q>0$. Completing the square for smooth compactly supported
wave functions gives the kinetic lower bound with this expression
integrated against $|\eta(X)|^2$. Dropping the nonnegative
regularization terms and letting $\varepsilon,\zeta$ tend to zero,
Fatou's lemma gives
\begin{equation}
 K[\eta]\ge
 \int\left(\frac{q^2}{8R^2}\sum_{j=1}^9\frac1{s_j}
                +\frac{968}{R^2}\right)|\eta(X)|^2\,\dd X.
 \label{eq:appE-Hardy}
\end{equation}
Approximating square-integrable wave functions with square-integrable
first derivatives by smooth compactly supported functions in the norm
that includes both the wave function and its first derivatives, and
using Fatou's lemma, extends the bound to the entire kinetic-energy
domain. For $d\ge9$, the set of coefficient matrices of rank less
than nine has zero volume. The bound ensures that the integral on
the right is finite for wave functions of finite kinetic energy.

The inverse-matrix contribution is related to the simplex-volume
inequality \cite{Lundholm}. The unregularized vector field can be
written as $-\nabla\log\phi$ with the auxiliary function
$\phi=(\det\mathsf W)^{-q/4}R^{-44}$, corresponding to the matrix
weight $\phi^2$ of ref.~\cite{MKS} for nine columns. The probability
density used throughout is $|\Omega(X)|^2$.

We now apply \eqref{eq:appE-Hardy} to the BPS wave function with the
weight $f=R^{-1/2}$. As in Appendix~\ref{app:inverse-details}, insert
smooth radial cutoffs and set $f_n=\chi_{a_n}\chi_{L_n}R^{-1/2}$,
where $a_n\to0$, $L_n\to\infty$ and the cutoffs vary only in
$(a_n,2a_n)$ and $(L_n,2L_n)$, respectively. Equation~\eqref{eq:inverse}
at $\nu=3$ gives $\ev{R^{-3}}<\infty$ and hence $\ev{R^{-1}}<\infty$.
Thus $f_n\Omega$ converges to $f\Omega$ in the square-integrable norm.
The squared-gradient errors from the two cutoffs are bounded by
\[
 C\int_{R<2a_n}\dd X\,R^{-3}|\Omega(X)|^2
\]
and by a constant times $L_n^{-3}$, respectively; both tend to zero.
Applying the supersymmetric multiplication identity to the difference
of two weights gives
\begin{equation*}
 h[(f_n-f_m)\Omega]
 =\frac12\ev{|\nabla f_n-\nabla f_m|^2}\longrightarrow0.
\end{equation*}
Closedness of the energy form then implies that $f\Omega$ has finite
energy form and $h[f\Omega]=\ev{R^{-3}}/8$. This does not require
$Hf\Omega$ itself to be square integrable.

At finite cutoff, using $V_B=g^2UR^4/4$ and the fermionic lower bound
$V_F(X)\ge-gC_NR\,I$ of \eqref{eq:yukawa}, we obtain
\begin{equation*}
 K[f_n\Omega]
 +\frac{g^2}{4}\ev{UR^4 f_n^2}
 \le h[f_n\Omega]+gC_N\ev{Rf_n^2},\qquad Rf_n^2\le1.
\end{equation*}
Apply \eqref{eq:appE-Hardy} to the kinetic term and remove the cutoffs
using Fatou's lemma on the nonnegative left-hand terms. The right-hand
side tends to $\ev{R^{-3}}/8+gC_N$, giving
\begin{align*}
 &\frac{q^2}{8}\ev{\frac1{R^3}\sum_{j=1}^9\frac1{s_j}}
 +\frac{g^2}{4}\ev{UR^3}+968\ev{R^{-3}}\\
 &\hspace{45pt}\le\frac18\ev{R^{-3}}+gC_N.
\end{align*}
Since $968-1/8=7743/8$, this is precisely \eqref{eq:spatial-mixed},
with all three terms on its left finite. The derivation does not
assume separate finiteness of the kinetic and potential energies of
$\Omega$.

\subsection{Joint constraints on spatial shape, extent and noncommutativity}

Cauchy--Schwarz applied separately to the first $k$ and last $9-k$
eigenvalues gives, with $\tau_k$ defined in \eqref{eq:spatial-tail},
\begin{equation}
 \sum_{j=1}^9\frac1{s_j}\ge\frac{k^2}{1-\tau_k}
 +\frac{(9-k)^2}{\tau_k}\ge\frac{(9-k)^2}{\tau_k}.
 \label{eq:appE-tail}
\end{equation}
The arithmetic--geometric mean inequality applied to the first two
terms of \eqref{eq:spatial-mixed} at each configuration gives
\[
 \frac{q^2}{8R^3}\sum_{j=1}^9\frac1{s_j}+\frac{g^2}{4}UR^3
 \ge\frac{gq}{2\sqrt2}\sqrt{U\sum_{j=1}^9\frac1{s_j}}.
\]
Taking expectations and using \eqref{eq:appE-tail} yields
\begin{equation}
 \ev{\sqrt{NU\sum_{j=1}^9 1/s_j}}\le\frac{16N\sqrt{N^2-1}}{q},
 \qquad
 \ev{\sqrt{NU/\tau_k}}\le\frac{16N\sqrt{N^2-1}}{(9-k)q}.
 \label{eq:appE-AMGM}
\end{equation}
For configurations satisfying $\tau_k\le\epsilon$ and $NU\ge u$,
with $\epsilon,u>0$, the quantity $\sqrt{NU/\tau_k}$ is at least
$\sqrt{u/\epsilon}$. Consequently,
\[
 \int_{\substack{\tau_k\le\epsilon\\NU\ge u}}\dd X\,|\Omega(X)|^2
 \le\sqrt{\frac\epsilon u}\ev{\sqrt{NU/\tau_k}},
\]
and \eqref{eq:appE-AMGM} gives the joint bound \eqref{eq:spatial-joint-U}.

For the joint bound involving the overall extent, note that
$k^2/(1-\tau_k)+(9-k)^2/\tau_k$ increases as $\tau_k$ decreases in
$0<\tau_k\le(9-k)/9$. Thus for $\tau_k\le\epsilon$ and
$R\le Lr_{3,N}$,
\[
 \frac{q^2}{8R^3}\sum_{j=1}^9\frac1{s_j}
 \ge\frac{q^2}{8L^3r_{3,N}^3}
 \left(\frac{k^2}{1-\epsilon}+\frac{(9-k)^2}{\epsilon}\right),
\]
where $0<\epsilon<(9-k)/9$ and $L>0$. Its expectation is at most
$gC_N$ by \eqref{eq:spatial-mixed}. Integrating over these configurations
as above and using $r_{3,N}^3=A_3/(gC_N)$ gives
\eqref{eq:spatial-joint-R}.

Suppose, for fixed $k$, that the probability of $\tau_k>\epsilon$
tends to zero for every fixed $\epsilon>0$. For example,
\begin{align*}
 \int_{NU\ge u}\dd X\,|\Omega(X)|^2
 \le{}&\int_{\tau_k>\epsilon}\dd X\,|\Omega(X)|^2\\
 &+\int_{\substack{\tau_k\le\epsilon\\NU\ge u}}\dd X\,|\Omega(X)|^2.
\end{align*}
The first term tends to zero by assumption, and the second is bounded
by \eqref{eq:spatial-joint-U}. The same splitting applies to
$R\le Lr_{3,N}$ using \eqref{eq:spatial-joint-R}. The factors
$N\sqrt{N^2-1}/q$ and $8A_3/q^2$ approach 1 and 81, respectively.
Taking $N\to\infty$ at fixed $u,L,\epsilon$ and then $\epsilon\to0$
shows that the probabilities of $NU\ge u$ and $R\le Lr_{3,N}$
both tend to zero for every fixed $u,L>0$.

To pass from probabilities to moments, fix $1\le p<\infty$ and split
 the expectation into the regions $NU\le u$ and $NU>u$. By
Cauchy--Schwarz,
\[
 \ev{(NU)^p}\le u^p+\ev{(NU)^{2p}}^{1/2}
 \left(\int_{NU>u}\dd X\,|\Omega(X)|^2\right)^{1/2}.
\]
The moment bound \eqref{eq:Lp} at the fixed exponent $2p$ bounds the
factor $\ev{(NU)^{2p}}^{1/2}$ for sufficiently large $N$.
The second term therefore vanishes as $N\to\infty$. Since $u>0$
can then be taken arbitrarily small, $\ev{(NU)^p}\to0$, completing
 the proof of \eqref{eq:spatial-necessary}.

The relation between overall extent and noncommutativity can also be
obtained without the spatial-shape estimate. Taking $\nu=3$ in
\eqref{eq:mixed} gives $g^2\ev{UR^3}/4\le gC_N$, or equivalently
\begin{equation}
 \ev{(NU)(R/r_{3,N})^3}\le\frac{4NC_N^2}{A_3}.
 \label{eq:appE-old-mixed}
\end{equation}
The upper bound approaches $1024/81$ at large $N$. Splitting the
expectation of $NU$ at $R=Lr_{3,N}$ gives
\begin{align*}
 \int_{R>Lr_{3,N}}\dd X\,NU|\Omega(X)|^2
 &\le\frac{4NC_N^2}{A_3L^3},\\
 \int_{R\le Lr_{3,N}}\dd X\,NU|\Omega(X)|^2
 &\le\ev{(NU)^2}^{1/2}
 \left(\int_{R\le Lr_{3,N}}\dd X\,|\Omega(X)|^2\right)^{1/2}.
\end{align*}
If $R/r_{3,N}$ grows without bound in probability, taking
$N\to\infty$ and then $L\to\infty$ gives $\ev{NU}\to0$. The higher
fixed moments then follow from the probability bound obtained from
this first moment and the splitting argument above. Thus the radial
estimate already links growth in extent to further suppression of
relative noncommutativity. The kinetic estimate of this appendix
supplies the new link to concentration in spatial directions.

\section{Supersymmetry and bounds on fluctuations}
\label{app:projected-poincare}

For a real invariant observable $F$ with a finite second moment,
its variance in a normalized BPS state $\Omega$ is
\[
 \ev{F^2}-\ev F^2=\norm{(F-\ev F)\Omega}^2.
\]
If the physical zero-energy state is unique, this is the squared norm
of the component of $F\Omega$ orthogonal to the zero-energy subspace.
More generally, let $P_0$ denote the orthogonal projection onto that
subspace. We first bound $\norm{(1-P_0)F\Omega}^2$ without assuming
uniqueness, and then specify when it equals the usual variance.

The supercharges map a rotation-invariant state $F\Omega$ into the
Spin(9) spinor sector. An energy bound in this sector will allow us
to control the projected fluctuation using derivatives of $F$.
We use the supersymmetric rotation identity \cite{HH} and extend the
spinor estimate in ref.~\cite[Appendix B]{MSY} to general color size.

\subsection{Energy bound in the spinor sector}

The generators $J_{ij}$ of transverse rotations can be expressed
as anticommutators with the supercharges. To reproduce both the
orbital and fermionic spin contributions to the rotation generator,
we use the following real $16\times16$ matrices and operators
for $i<j$:
\begin{align}
 \mathsf A_a^{ij}
 &\coloneqq\frac{X_{ai}\Gamma_j-X_{aj}\Gamma_i}{16}
   +\frac1{56}\sum_{k\ne i,j}X_{ak}\Gamma_k\Gamma_i\Gamma_j,
 \notag\\
 \tau_\alpha^{ij}
 &\coloneqq\sum_{a,\beta}(\mathsf A_a^{ij})_{\alpha\beta}\theta_\beta^a,
 \label{eq:appB-tau}\\
 J_{ij}
 &\coloneqq-\ii\sum_a(X_{ai}\partial_{aj}-X_{aj}\partial_{ai})
   -\frac{\ii}{4}\sum_{a,\beta,\gamma}\theta_\beta^a
                 (\Gamma_i\Gamma_j)_{\beta\gamma}\theta_\gamma^a.
 \label{eq:appB-J}
\end{align}
On the common physical core,
\begin{equation}
 \sum_{\alpha=1}^{16}\{q_\alpha,\tau_\alpha^{ij}\}=J_{ij}.
 \label{eq:appB-Qexact}
\end{equation}
For completeness, the derivative terms are fixed by the following
identities, where $\Tr_{16}$ is the ordinary trace
on the sixteen-dimensional spinor index space:
\[
 \Tr_{16}(\Gamma_k\mathsf A_a^{ij})
 =X_{ai}\delta_{kj}-X_{aj}\delta_{ki},\qquad
 \sum_k\Gamma_k\partial_{ak}\mathsf A_a^{ij}
 =\left(\frac2{16}+\frac7{56}\right)\Gamma_i\Gamma_j
 =\frac14\Gamma_i\Gamma_j.
\]
The interaction term in the anticommutator is proportional to
\[
 \sum_{a,b,c,k,l}f^{abc}X_{bk}X_{cl}
   \Tr_{16}\!\left[(\Gamma_{kl})^T
                                      \mathsf A_a^{ij}\right].
\]
Each trace vanishes before the color indices are contracted: after repeated
indices are removed, it is the trace of an odd product of at most five
distinct gamma matrices, and conjugation by an unused gamma matrix reverses
its sign.  Thus no identity specific to $SU(2)$ is involved.

Set
\[
 \tau_\alpha\coloneqq\tau_\alpha^{12}+\tau_\alpha^{34}
                       +\tau_\alpha^{56}+\tau_\alpha^{78},\qquad
 \Xi_\alpha\coloneqq\sum_{a,k,\beta}X_{ak}
                  (\Gamma_k)_{\alpha\beta}\theta_\beta^a,
 \qquad S_9\coloneqq\sum_aX_{a9}^2.
\]
The canonical anticommutation relations and gamma-matrix traces give
\begin{align}
 \sum_\alpha(\tau_\alpha^{ij})^2
 &=\sum_a\left[\frac{X_{ai}^2+X_{aj}^2}{32}
             +\frac1{392}\sum_{k\ne i,j}X_{ak}^2\right],
 \notag\\
 \sum_\alpha\tau_\alpha^2
 &=\frac{61R^2-45S_9}{1568},\qquad
 \sum_\alpha\Xi_\alpha^2=8R^2.
 \label{eq:appB-sums}
\end{align}
In the first sum, we use the canonical anticommutation relation
\(\bigl(\sum_{a,\beta}v_{a\beta}\theta_\beta^a\bigr)^2
=\tfrac12\sum_{a,\beta}v_{a\beta}^2 I\) for real \(v\).
The traces of the products of the transposed coefficient matrices
from distinct planes vanish, which gives the second line.  All traces here
are over the 16-dimensional gamma-matrix space; there is no trace over
the fermionic Hilbert space and no extra factor depending on its dimension.

Let $\HH_{16}$ denote the physical subspace consisting of copies of the
Spin(9) spinor representation $\mathbf{16}$. For a state
$\eta$ in $\HH_{16}$ with finite energy form (that is,
$\eta\in\Dom(H^{1/2})$), we will show that
\begin{equation}
 h[\eta]\ge c_{16}\norm{R^{-1}\eta}^2,
 \qquad c_{16}\coloneqq\frac{9-2\sqrt{14}}4,
 \qquad \kappa_{16}\coloneqq c_{16}^{-1}
                 =\frac{36+8\sqrt{14}}{25}.
 \label{eq:appB-hardy}
\end{equation}
The constants are independent of $N$ and $g$.
To derive the bound, note that $H$ and $R$ commute with Spin(9).
They act trivially on the spinor representation factor (though not
necessarily on the multiplicity space). It is therefore sufficient to
prove the bound for highest-weight vectors with arbitrary multiplicity.
With $J\coloneqq J_{12}+J_{34}+J_{56}+J_{78}$, these satisfy $J\eta=2\eta$.
The decomposition
\[
 \mathbf{16}^*\otimes\mathbf{16}
 =\mathbf1\oplus\mathbf9\oplus\mathbf{36}
                      \oplus\mathbf{84}\oplus\mathbf{126}
\]
contains no symmetric traceless rank-two tensor $\mathbf{44}$.
Only the rotational scalar part of the quadratic coordinate tensor
contributes between spinor states. Consequently
$\langle\eta,S_9R^{-4}\eta\rangle
=\norm{R^{-1}\eta}^2/9$.
Equation~\eqref{eq:appB-sums} then gives
\[
 \sum_\alpha\norm{R^{-2}\tau_\alpha\eta}^2
 =\frac1{28}\norm{R^{-1}\eta}^2,
 \qquad
 \sum_\alpha\norm{R^{-2}\Xi_\alpha\eta}^2=8\norm{R^{-1}\eta}^2.
\]
Summing~\eqref{eq:appB-Qexact} over the four rotation planes and
using $J\eta=2\eta$, we obtain
\[
 2\norm{R^{-1}\eta}^2=\langle\eta,R^{-2}J\eta\rangle
 =\sum_\alpha\langle\eta,R^{-2}\{q_\alpha,\tau_\alpha\}\eta\rangle.
\]
Moving the supercharge in the first term of the anticommutator
across the inner product introduces the commutator
$[q_\alpha,R^{-2}]=2\ii R^{-4}\Xi_\alpha$. Since
$\tau_\alpha$ is Hermitian and commutes with $R$, the
Cauchy--Schwarz inequality gives
\begin{align*}
 2\norm{R^{-1}\eta}^2
 &\le 2\left(\sum_\alpha\norm{q_\alpha\eta}^2\right)^{1/2}
       \left(\sum_\alpha\norm{R^{-2}\tau_\alpha\eta}^2\right)^{1/2}\\
 &\quad+2\left(\sum_\alpha\norm{R^{-2}\Xi_\alpha\eta}^2\right)^{1/2}
       \left(\sum_\alpha\norm{R^{-2}\tau_\alpha\eta}^2\right)^{1/2}.
\end{align*}
Using $\sum_\alpha\norm{q_\alpha\eta}^2=16h[\eta]$ and the two
norm identities above yields
\begin{equation*}
 2\norm{R^{-1}\eta}^2
 \le\frac4{\sqrt7}\sqrt{h[\eta]}\norm{R^{-1}\eta}
     +2\sqrt{\frac27}\norm{R^{-1}\eta}^2.
\end{equation*}
For a nonzero state, rearranging this inequality and squaring gives
\[
 h[\eta]\ge\frac74\left(1-\sqrt{\frac27}\right)^2
                \norm{R^{-1}\eta}^2
 =\frac{9-2\sqrt{14}}4\norm{R^{-1}\eta}^2,
\]
which is~\eqref{eq:appB-hardy}. The zero state satisfies it trivially.

Initially these calculations are made on the physical core with support
away from $R=0$.  Multiplication of a smooth compactly supported vector by
an inner radial cutoff produces a supercharge commutator whose
squared norm is $O(\epsilon^{D-2})$. This term vanishes as the
cutoff is removed, and the cutoff also converges to one in the
Hilbert-space norm on the state and on its supercharge image.
The scalar cutoff commutes with the potential and with the gauge
and Spin(9) projections. Thus any state in the required
representation with finite energy form can be approximated by these
trial states, both in Hilbert-space norm and in the energy-form norm.
Applying the estimate to differences of the approximating states shows
that the sequence obtained by multiplying them by $R^{-1}$ is Cauchy
in Hilbert-space norm. The closedness of multiplication by $R^{-1}$
then extends the estimate to the stated domain.
The bound is weighted by $R^{-2}$; it does not require a positive
constant spectral gap. We do not need the optimal coefficient.

\subsection{Bounds on fluctuations of invariant observables}

Normalizable physical zero-energy states
are Spin(9) invariant~\cite{HH,SS}; this established fact is used in the
rest of the appendix.

For a normalized physical zero-energy state $\Omega$, consider a
real gauge- and Spin(9)-invariant function $F$.  If $F$ is smooth and
bounded with compactly supported gradient, then
\begin{equation}
 \norm{(1-P_0)F\Omega}^2
 \le\CP\ev{R^2|\nabla F|^2},
 \qquad \CP\coloneqq\frac{\kappa_{16}}2
           =\frac{18+4\sqrt{14}}{25}.
 \label{eq:appB-poincare}
\end{equation}
The inequality also holds in the limit of approximating functions
$F_n$ if both $F_n\Omega\to F\Omega$ in Hilbert-space norm and
$\ev{R^2|\nabla F_n-\nabla F|^2}\to0$.

To derive the inequality, define $Y_\alpha\coloneqq\ii[q_\alpha,F]$.
The canonical anticommutation relations give
\begin{equation}
 Y_\alpha=\sum_{a,i,\beta}(\partial_{ai}F)(\Gamma_i)_{\alpha\beta}\theta_\beta^a,
 \qquad \sum_\alpha Y_\alpha^2=8|\nabla F|^2.
 \label{eq:appB-CARF}
\end{equation}
Multiplying $F\Omega$ by the outer radial cutoff of
Appendix~\ref{app:inverse-details} gives smooth physical test states
converging in norm, with their supercharge images converging to
$[q_\alpha,F]\Omega$ since $F$ is bounded and $\nabla F$ has compact
support. Closedness of $q_\alpha$ thus places $F\Omega$ in its domain.
As $F\Omega$ is a singlet,
$\xi_\alpha\coloneqq q_\alpha F\Omega=[q_\alpha,F]\Omega$ lies in $\HH_{16}$.
To use the spinor energy bound to estimate an inverse of $H$,
consider a spinor state $\xi\in\HH_{16}$ for which $R\xi$ is square
integrable. For $\epsilon>0$, set
$\chi=(H+\epsilon)^{-1}\xi$. Since $H$ commutes with Spin(9),
$\chi$ belongs to the same spinor sector and has finite energy form.
Equation~\eqref{eq:appB-hardy} gives
\begin{align*}
 c_{16}\norm{R^{-1}\chi}^2
 &\le h[\chi]\le\langle\chi,(H+\epsilon)\chi\rangle
 =\langle\chi,\xi\rangle,\\
 \langle\chi,\xi\rangle
 &=\langle R^{-1}\chi,R\xi\rangle
 \le\norm{R^{-1}\chi}\norm{R\xi}.
\end{align*}
The inner product here is real and nonnegative by the definition of
$\chi$. Combining the two inequalities gives
\[
 \langle\xi,(H+\epsilon)^{-1}\xi\rangle
 \le\kappa_{16}\norm{R\xi}^2.
\]
Taking $\epsilon\downarrow0$ and using $q_\alpha^2=H$ in the spectral
calculus gives
\[
 \norm{(1-P_0)F\Omega}^2
 =\langle\xi_\alpha,H^{-1}\xi_\alpha\rangle
 \le\kappa_{16}\norm{R\xi_\alpha}^2.
\]
Here the inverse is taken on the positive-energy spectrum, in the
sense of the limit of the regularized matrix element above. Averaging
over the 16 values of $\alpha$ and using~\eqref{eq:appB-CARF} proves
\eqref{eq:appB-poincare}. This is a projected Poincar\'e inequality.

For the usual variance $\ev{F^2}-\ev F^2$, the relevant wave function
is $(F-\ev F)\Omega$. It equals $(1-P_0)F\Omega$ if
\begin{equation}
 P_0F\Omega=\ev F\,\Omega.
 \label{eq:appB-projection}
\end{equation}
Under this condition,~\eqref{eq:appB-poincare} gives
\[
 \ev{F^2}-\ev F^2\le\CP\ev{R^2|\nabla F|^2}.
\]
Uniqueness, $\ker H=\mathbb C\Omega$, is sufficient. Without uniqueness,
\eqref{eq:appB-projection} must hold for every insertion used in
a particular argument.  Its validity for $F$ does not imply its validity
for a nonlinear function of $F$.

\section{Bounds on fluctuations of the matrix extent}
\label{app:log-radius}

Under the projection condition~\eqref{eq:appB-projection}, the variance
bound of Appendix~\ref{app:projected-poincare} controls fluctuations of
an invariant observable $F$ in terms of $\ev{R^2|\nabla F|^2}$.
For $F=\log(R/R_0)$, with an arbitrary fixed $R_0>0$, one has
$R^2|\nabla F|^2=1$ away from the origin. Thus the logarithm is a natural
observable for obtaining a bound on relative fluctuations of the matrix
extent, with a constant independent of $N$ and $g$. To justify using this
unbounded observable, we first apply the inequality to bounded smooth
approximations, imposing the projection condition on each of them.

Write $Y\coloneqq\log(R/R_0)$. Let $h_L(y)$ be smooth bounded
approximations to $y$, with $h_L(0)=0$, $|h_L'|\le1$, $h_L(y)=y$ on
$[-L,L]$, and constant tails. Assume~\eqref{eq:appB-projection} for all
$h_L(Y)$; uniqueness of the physical zero-energy state is a sufficient
condition. Since $R|\nabla h_L(Y)|\le1$, eq.~\eqref{eq:appB-poincare}
gives $\Var(h_L(Y))\le\CP$.
Since $Y$ is finite with probability one, choose a finite $y_0\ge0$
with $\int_{|Y|\le y_0}\dd X\,|\Omega(X)|^2\ge1/2$.
For $L\ge y_0$, $|h_L(Y)|\le y_0$ in this region. If
$|\ev{h_L(Y)}|>y_0$, the contribution of this region to the variance
is at least $(|\ev{h_L(Y)}|-y_0)^2/2$. Consequently, in either case,
\[
 |\ev{h_L(Y)}|\le y_0+\sqrt{2\CP},\qquad
 \ev{h_L(Y)^2}\le\CP+(y_0+\sqrt{2\CP})^2.
\]
As $L\to\infty$, $h_L(Y)\to Y$ pointwise almost everywhere, so Fatou's
lemma gives $\ev{Y^2}<\infty$. Since $|h_L(Y)|\le |Y|$, the
approximations then converge in mean square by dominated convergence.
Taking the limit of the variance bound gives
\begin{equation}
 \Var\log(R/R_0)\le\CP.
 \label{eq:appD-logvar}
\end{equation}
In particular, the finiteness of the logarithmic second moment is a
consequence of the bound, not an assumption.

To see what this means for configurations at widely separated radial
scales, consider $0<r_-<r_+<\infty$ and write
\[
 q_-=\int_{R\le r_-}\dd X\,|\Omega(X)|^2,\qquad
 q_+=\int_{R\ge r_+}\dd X\,|\Omega(X)|^2.
\]
If both probabilities are positive, let $\mu_-$ and $\mu_+$ denote the
conditional means of $Y$ in the two regions. They satisfy
$\mu_-\le\log(r_-/R_0)$ and $\mu_+\ge\log(r_+/R_0)$.
Restricting the variance integral to these two regions and applying
Cauchy--Schwarz in each gives
\[
 \Var(Y)\ge q_-(\mu_--\ev Y)^2+q_+(\mu_+-\ev Y)^2
 \ge\frac{q_-q_+}{q_-+q_+}(\mu_+-\mu_-)^2.
\]
The second inequality follows by minimizing the quadratic expression
with respect to $\ev Y$ over the whole real line, so no assumption
about the location of the mean is needed. Together with
\eqref{eq:appD-logvar}, this implies
\begin{equation}
 \frac{q_-q_+}{q_-+q_+}
        \left(\log\frac{r_+}{r_-}\right)^2\le\CP,
 \label{eq:appD-two-scales}
\end{equation}
for $q_-+q_+>0$; if either probability is zero the bound is trivial.
Thus two radial ranges with appreciable probabilities cannot be
arbitrarily far apart in logarithmic radius.

We next ask whether the first positive and inverse moments of $R$
can both be controlled relative to a common radius. This involves
exponentials of the logarithm and requires the projection condition
for additional functions. In addition to all the bounded approximations
above, require~\eqref{eq:appB-projection} for
\begin{equation}
 F\coloneqq\exp\!\left[\frac{t}{2^{j+1}}h_L(Y)\right],
 \qquad L>0,\quad j=0,1,2,\ldots,
 \label{eq:appD-all-insertions}
\end{equation}
for every value of $t$ used below.
The conditions on $h_L(Y)$ alone do not imply these exponential
conditions. Subtracting the mean in the exponent only multiplies $F$
by a scalar and therefore does not change the condition. To obtain all
the following estimates at once, one may impose the projection condition
for the functions in~\eqref{eq:appD-all-insertions} for all
$|t|<2/\sqrt{\CP}$.

Define
$M_L(t)\coloneqq\ev{\exp[t(h_L(Y)-\ev{h_L(Y)})]}$.
Apply the variance bound to
$F=\exp[\tfrac{t}{2}(h_L(Y)-\ev{h_L(Y)})]$.
Its variance is $M_L(t)-M_L(t/2)^2$, while
\[
 R^2|\nabla F|^2=\frac{t^2}{4}F^2 R^2|\nabla h_L(Y)|^2
 \le\frac{t^2}{4}F^2.
\]
The variance bound therefore gives
\begin{equation*}
 \left(1-\frac{\CP t^2}{4}\right)M_L(t)
 \le M_L(t/2)^2.
\end{equation*}
For $|t|<2/\sqrt{\CP}$, all factors on the left are positive at
successive half-arguments. Iterating the inequality, as in the
exponential-moment method of ref.~\cite{BL}, gives
\[
 M_L(t)\le M_L(t/2^n)^{2^n}
 \prod_{j=0}^{n-1}\left(1-\frac{\CP t^2}{4^{j+1}}\right)^{-2^j}.
\]
For fixed $L$ the centered observable is bounded and has zero mean,
so $M_L(s)=1+O(s^2)$ as $s\to0$. The terminal factor tends to 1 as
$n\to\infty$. By convexity of $-\log(1-x)$, for
$0\le x<1$ one has $-\log(1-x/4^j)\le-4^{-j}\log(1-x)$.
With $x=\CP t^2/4$ and $\sum_{j\ge0}2^{-j}=2$, this bounds the infinite
product independently of $L$. The mean-square convergence of $h_L(Y)$
and Fatou's lemma then give
\begin{equation}
 \begin{aligned}
 \ev{e^{t(Y-\ev Y)}}
 &\le\prod_{j=0}^{\infty}
   \left(1-\frac{\CP t^2}{4^{j+1}}\right)^{-2^j}\\
 &\le\left(1-\frac{\CP t^2}{4}\right)^{-2},
 \qquad |t|<\frac2{\sqrt{\CP}}=\sqrt{14}-2.
 \end{aligned}
 \label{eq:appD-mgf}
\end{equation}
No exponential integrability is assumed before this argument.

In particular, $t=\pm1$ is admissible. With the geometric mean radius
$\rho\coloneqq R_0\exp\ev{\log(R/R_0)}$ and
$B\coloneqq(1-\CP/4)^{-2}$, eq.~\eqref{eq:appD-mgf} and Jensen's
inequality give
\begin{equation}
 \rho\le\ev R\le B\rho,\qquad
 \rho^{-1}\le\ev{R^{-1}}\le B\rho^{-1},\qquad
 1\le\ev R\,\ev{R^{-1}}\le B^2.
 \label{eq:appD-first-moments}
\end{equation}
Under the full set of projection conditions stated above, the arithmetic
mean radius and the harmonic mean radius $1/\ev{R^{-1}}$ thus lie within
a fixed factor of the geometric mean radius $\rho$, independent of $N$
and $g$. These bounds control the spread in radial scales separately
from the location of the distribution: determining the absolute scale
$\rho$ remains a further physical question. In this way they supplement
the main inverse-radius estimates.

\bibliographystyle{JHEP}
\bibliography{BFSS_References_v12}

\providecommand{\href}[2]{#2}\begingroup\raggedright\begin{thebibliography}{10}

\bibitem{BFSS}
T.~Banks, W.~Fischler, S.H.~Shenker and L.~Susskind, \emph{{M} theory as a
  matrix model: A conjecture},
  \href{https://doi.org/10.1103/PhysRevD.55.5112}{\emph{Phys. Rev. D}
  {\bfseries 55} (1997) 5112}
  [\href{https://arxiv.org/abs/hep-th/9610043}{{\ttfamily hep-th/9610043}}].

\bibitem{SusskindDLCQ}
L.~Susskind, \emph{Another conjecture about {M(atrix)} theory},
  \href{https://arxiv.org/abs/hep-th/9704080}{{\ttfamily hep-th/9704080}}.

\bibitem{SenDLCQ}
A.~Sen, \emph{{D0} branes on {$T^n$} and {Matrix} theory},
  \href{https://doi.org/10.4310/ATMP.1998.v2.n1.a2}{\emph{Adv. Theor. Math.
  Phys.} {\bfseries 2} (1998) 51}
  [\href{https://arxiv.org/abs/hep-th/9709220}{{\ttfamily hep-th/9709220}}].

\bibitem{SeibergDLCQ}
N.~Seiberg, \emph{Why is the matrix model correct?},
  \href{https://doi.org/10.1103/PhysRevLett.79.3577}{\emph{Phys. Rev. Lett.}
  {\bfseries 79} (1997) 3577}
  [\href{https://arxiv.org/abs/hep-th/9710009}{{\ttfamily hep-th/9710009}}].

\bibitem{SS}
S.~Sethi and M.~Stern, \emph{Invariance theorems for supersymmetric
  {Yang--Mills} theories},
  \href{https://doi.org/10.4310/atmp.2000.v4.n2.a8}{\emph{Adv. Theor. Math.
  Phys.} {\bfseries 4} (2000) 487}
  [\href{https://arxiv.org/abs/hep-th/0001189}{{\ttfamily hep-th/0001189}}].

\bibitem{HH}
D.~Hasler and J.~Hoppe, \emph{Zero energy states of reduced super {Yang--Mills}
  theories in {$d+1=4,6$} and {$10$} dimensions are necessarily
  {$\operatorname{Spin}(d)$} invariant},
  \href{https://arxiv.org/abs/hep-th/0211226}{{\ttfamily hep-th/0211226}}.

\bibitem{FGHHY}
J.~Fr{\"o}hlich, G.M.~Graf, D.~Hasler, J.~Hoppe and S.-T.~Yau, \emph{Asymptotic
  form of zero energy wave functions in supersymmetric matrix models},
  \href{https://doi.org/10.1016/S0550-3213(99)00649-5}{\emph{Nucl. Phys. B}
  {\bfseries 567} (2000) 231}
  [\href{https://arxiv.org/abs/hep-th/9904182}{{\ttfamily hep-th/9904182}}].

\bibitem{HHasy}
D.~Hasler and J.~Hoppe, \emph{Asymptotic factorisation of the ground-state for
  {SU(N)}-invariant supersymmetric matrix-models},
  \href{https://arxiv.org/abs/hep-th/0206043}{{\ttfamily hep-th/0206043}}.

\bibitem{LY}
Y.-H.~Lin and X.~Yin, \emph{On the ground state wave function of matrix
  theory}, \href{https://doi.org/10.1007/JHEP11(2015)027}{\emph{JHEP}
  {\bfseries 11} (2015) 027} [\href{https://arxiv.org/abs/1402.0055}{{\ttfamily
  1402.0055}}].

\bibitem{P}
J.~Polchinski, \emph{{M}-theory and the light cone},
  \href{https://doi.org/10.1143/PTPS.134.158}{\emph{Prog. Theor. Phys. Suppl.}
  {\bfseries 134} (1999) 158}
  [\href{https://arxiv.org/abs/hep-th/9903165}{{\ttfamily hep-th/9903165}}].

\bibitem{L}
H.W.~Lin, \emph{Bootstrap bounds on {D0}-brane quantum mechanics},
  \href{https://doi.org/10.1007/JHEP06(2023)038}{\emph{JHEP} {\bfseries 06}
  (2023) 038} [\href{https://arxiv.org/abs/2302.04416}{{\ttfamily
  2302.04416}}].

\bibitem{LZ}
H.W.~Lin and Z.~Zheng, \emph{Bootstrapping ground state correlators in matrix
  theory, {Part I}}, \href{https://doi.org/10.1007/JHEP01(2025)190}{\emph{JHEP}
  {\bfseries 01} (2025) 190}
  [\href{https://arxiv.org/abs/2410.14647}{{\ttfamily 2410.14647}}].

\bibitem{MSY}
A.~Michel, J.~Scheinpflug and X.~Yin, \emph{{99.9...\% of the SU(2) BFSS ground
  state}},  \href{https://arxiv.org/abs/2611.09837}{{\ttfamily 2611.09837}}.

\bibitem{BLin}
K.~Bajaj and H.W.~Lin, \emph{Solving the {SU(2) BFSS} model},
  \href{https://arxiv.org/abs/2611.14512}{{\ttfamily 2611.14512}}.

\bibitem{FS}
R.L.~Frank and R.~Seiringer, \emph{Non-linear ground state representations and
  sharp {Hardy} inequalities},
  \href{https://doi.org/10.1016/j.jfa.2008.05.015}{\emph{J. Funct. Anal.}
  {\bfseries 255} (2008) 3407}
  [\href{https://arxiv.org/abs/0803.0503}{{\ttfamily 0803.0503}}].

\bibitem{GXYYZ}
J.~Ge, S.~Xu, H.~You and Y.~Zhou, \emph{{DDVV}-type inequality for {Hermitian}
  matrices}, \href{https://doi.org/10.1016/j.laa.2017.04.028}{\emph{Linear
  Algebra Appl.} {\bfseries 529} (2017) 133}
  [\href{https://arxiv.org/abs/1610.05020}{{\ttfamily 1610.05020}}].

\bibitem{dWHN}
B.~de~Wit, J.~Hoppe and H.~Nicolai, \emph{On the quantum mechanics of
  supermembranes},
  \href{https://doi.org/10.1016/0550-3213(88)90116-2}{\emph{Nucl. Phys. B}
  {\bfseries 305} (1988) 545}.

\bibitem{dWLN}
B.~de~Wit, M.~L{\"u}scher and H.~Nicolai, \emph{The supermembrane is unstable},
  \href{https://doi.org/10.1016/0550-3213(89)90214-9}{\emph{Nucl. Phys. B}
  {\bfseries 320} (1989) 135}.

\bibitem{BGR}
L.~Boulton, M.P.~Garc{\'i}a~del Moral and A.~Restuccia, \emph{Existence of a
  supersymmetric massless ground state of the {SU(N)} matrix model globally on
  its valleys}, \href{https://doi.org/10.1007/JHEP05(2021)281}{\emph{JHEP}
  {\bfseries 05} (2021) 281}
  [\href{https://arxiv.org/abs/2102.00886}{{\ttfamily 2102.00886}}].

\bibitem{HMSoft}
A.~Herderschee and J.~Maldacena, \emph{Soft theorems in matrix theory},
  \href{https://doi.org/10.1007/JHEP11(2024)052}{\emph{JHEP} {\bfseries 11}
  (2024) 052} [\href{https://arxiv.org/abs/2312.15111}{{\ttfamily
  2312.15111}}].

\bibitem{Chernoff77}
P.R.~Chernoff, \emph{{Schr\"odinger} and {Dirac} operators with singular
  potentials and hyperbolic equations},
  \href{https://doi.org/10.2140/pjm.1977.72.361}{\emph{Pacific J. Math.}
  {\bfseries 72} (1977) 361}.

\bibitem{vCS}
J.G.~van~der Corput and G.~Schaake, \emph{Ungleichungen f{\"u}r {Polynome} und
  trigonometrische {Polynome}}, {\emph{Compositio Math.} {\bfseries 2} (1935)
  321}.

\bibitem{vCSerr}
J.G.~van~der Corput and G.~Schaake, \emph{Berichtigung zu: {Ungleichungen}
  f{\"u}r {Polynome} und trigonometrische {Polynome}}, {\emph{Compositio Math.}
  {\bfseries 3} (1936) 128}.

\bibitem{Aronszajn}
N.~Aronszajn, \emph{A unique continuation theorem for solutions of elliptic
  partial differential equations or inequalities of second order}, {\emph{J.
  Math. Pures Appl.} {\bfseries 36} (1957) 235}.

\bibitem{Lundholm}
D.~Lundholm, \emph{Geometric extensions of many-particle {Hardy} inequalities},
  \href{https://doi.org/10.1088/1751-8113/48/17/175203}{\emph{J. Phys. A}
  {\bfseries 48} (2015) 175203}
  [\href{https://arxiv.org/abs/1101.2653}{{\ttfamily 1101.2653}}].

\bibitem{MKS}
T.~Matsuda, F.~Komaki and W.E.~Strawderman, \emph{Double shrinkage priors for a
  normal mean matrix}, \href{https://doi.org/10.1214/24-BA1457}{\emph{Bayesian
  Anal.} {\bfseries 21} (2026) 1}
  [\href{https://arxiv.org/abs/2311.13137}{{\ttfamily 2311.13137}}].

\bibitem{BL}
S.G.~Bobkov and M.~Ledoux, \emph{{Poincar\'e}'s inequalities and {Talagrand}'s
  concentration phenomenon for the exponential distribution},
  \href{https://doi.org/10.1007/s004400050090}{\emph{Probab. Theory Relat.
  Fields} {\bfseries 107} (1997) 383}.

\end{thebibliography}\endgroup
\end{document}